\documentclass[reprint,amsmath,amssymb,aps,prl]{revtex4-2}
\usepackage[utf8]{inputenc}
\usepackage{color}
\usepackage{amsmath}
\usepackage{graphicx}
\usepackage{xurl}
\usepackage[bookmarks=false,
 breaklinks=true,hypertexnames=false,pdfborder={0 0 1},backref=false,colorlinks=true]
 {hyperref}

\makeatletter

\usepackage{zref-xr}
\zxrsetup{toltxlabel}
\usepackage{orcidlink}
\usepackage{dcolumn}
\usepackage{bm}
\usepackage{bbm}
\usepackage{caption}
\usepackage{ragged2e}
\DeclareCaptionJustification{justified}{\justifying}

\ifdefined\showcaptionsetup
 
\fi

\ifdefined\showcaptionsetup
 
\fi
\ifdefined\showcaptionsetup
 \PassOptionsToPackage{caption=false}{subfig}
\fi
\usepackage{subfig}
\makeatother

\begin{document}
\title{Fundamental Limits of Quantum Metrology Beyond Fixed Causal Order}
\author{Wenjie Wei$^{1}$\orcidlink{0009-0001-2435-0852}, Yutong Li$^{1}$\orcidlink{0009-0003-9365-6913},
and Shengshi Pang$^{1,2}$\orcidlink{0000-0002-6351-539X}}
\email{pangshsh@mail.sysu.edu.cn}

\affiliation{$^{1}$School of Physics, Sun Yat-sen University, Guangzhou, Guangdong
510275, China}
\affiliation{$^{2}$Hefei National Laboratory, University of Science and Technology
of China, Hefei 230088, China}
\date{\today}
\begin{abstract}
Quantum metrology with indefinite causal order (ICO) has attracted
intense interest due to its potential to surpass the limitations of
conventional fixed-order strategies. A key open question is whether
ICO can fundamentally enhance asymptotic precision scaling. In this
work, we bridge this gap for the estimation of a single parameter
encoded in $N$ identical uses of a finite-dimensional quantum channel.
We first establish a universal Heisenberg-scaling upper bound for
the full general ICO process-matrix class and show that for unitary
channels its optimal quantum Fisher information (QFI) coincides exactly
with that of parallel strategies. For noisy channels, a structurally
refined bound shows that channels restricted to the standard quantum
limit (SQL) under parallel strategies remain SQL-limited under general
ICO strategies. Most significantly, an asymptotically
tight (AT) bound is derived to close the remaining possibility of
an asymptotic ICO advantage by showing that general ICO and optimal
parallel strategies have exactly the same leading QFI coefficient
in both the SQL and Heisenberg regimes. For the operationally motivated
class of quantum circuits with quantum control of causal order, we
further obtain an iterative constraint on finite-query precision
whose asymptotic limit agrees with that of the AT bound. Our results
clarify the ultimate role of indefinite causality as a metrological
resource for quantum channel estimation.
\end{abstract}
\maketitle
\textit{Introduction---}Quantum metrology seeks to determine the
ultimate precision in parameter estimation~\citep{escher2011GeneralFrameworkEstimating,demkowicz-dobrzanski2012ElusiveHeisenbergLimit}
by exploiting nonclassical resources such as entanglement~\citep{giovannetti2006QuantumMetrology},
squeezing~\citep{maccone2020SqueezingMetrologyUnified}, and coherence~\citep{braun2018QuantumenhancedMeasurementsEntanglement},
with applications from optical atomic clocks and quantum imaging~\citep{colombo2022EntanglementenhancedOpticalAtomic,genovese2016RealApplicationsQuantum}
to phase estimation~\citep{polino2020PhotonicQuantumMetrology},
quantum thermometry~\citep{mehboudi2019ThermometryQuantumRegime},
and continuous-variable metrology~\citep{fadel2025QuantumMetrologyContinuousvariable}.
It also admits different statistical formulations.
The local frequentist approach characterizes the precision through
the quantum Fisher information (QFI) and quantum Cramér--Rao bound~\citep{paris2009QUANTUMESTIMATIONQUANTUM},
whereas the Bayesian approach optimizes an average loss function over
a prior distribution~\citep{bavaresco2024DesigningOptimalProtocolsa}.
For a single parameter encoded in $N$ uses of a finite-dimensional
quantum channel, a key question is how the attainable QFI scales with
$N$. For unitary channels, parallel strategies with $N$ entangled
probes can attain the Heisenberg limit (HL) with uncertainty scaling
as $1/N$, surpassing the $1/\sqrt{N}$ standard quantum limit (SQL)
of independent probes.

Noise can fundamentally alter this scaling and reduce the precision
of entangled parallel schemes from the HL to the SQL. Adaptive strategies
involving intermediate operations between channel uses have been developed
to mitigate this limitation~\citep{demkowicz-dobrzanski2014UsingEntanglementNoise},
which can outperform parallel schemes at finite $N$ for specific
nonunitary channels~\citep{liu2023OptimalStrategiesQuantum} but
offer no advantage in the asymptotic limit~\citep{kurdzialek2023UsingAdaptivenessCausal,zhou2021AsymptoticTheoryQuantum}
or improvement over optimal parallel strategies for unitary channels~\citep{giovannetti2006QuantumMetrology,demkowicz-dobrzanski2014UsingEntanglementNoise}.
These conventional strategies nevertheless assume a fixed causal structure,
with operations arranged in a definite temporal order that can be
represented by a directed acyclic graph~\citep{barrett2020QuantumCausalModels}.

To go beyond fixed-order strategies, indefinite causal
order (ICO) has been introduced via the process-matrix formalism~\citep{oreshkov2012QuantumCorrelationsNo,vilasini2024FundamentalLimitsRealising,delahamette2025IndefiniteCausalOrder,costa2026IndefiniteQuantumCausality},
where quantum processes are not compatible with any predefined global
causal order. A canonical example ~\citep{araujo2015WitnessingCausalNonseparability,ognyanoreshkov2016CausalCausallySeparable}
is the quantum switch, where a control system coherently determines
the order of multiple operations, creating a quantum superposition
of causal paths~\citep{chiribella2013QuantumComputationsDefinite,goswami2018IndefiniteCausalOrder}.
ICO processes have been shown to offer advantages in various quantum
tasks, including channel discrimination~\citep{bavaresco2021StrictHierarchyParallel},
quantum computation~\citep{araujo2014ComputationalAdvantageQuantumControlled,taddei2021ComputationalAdvantageQuantum,abbott2025ClassicalQuantumQuery},
and communication~\citep{chiribella2021IndefiniteCausalOrder,chiribella2021QuantumClassicalData,wu2025GeneralCommunicationEnhancement}.
This has sparked intense interest in the potential of ICO in quantum
metrology~\citep{mukhopadhyay2018SuperpositionCausalOrder,chapeau-blondeau2021NoisyQuantumMetrology,liu2024FullyOptimizedQuantumMetrology,goldberg2023EvadingNoiseMultiparameter,chapeau-blondeau2022IndefiniteCausalOrder,frey2019IndefiniteCausalOrder,zhou2024StrictHierarchyOptimal}.

For certain noisy channels, it has been known that ICO strategies
can strictly outperform all fixed-order schemes~\citep{liu2023OptimalStrategiesQuantum,mothe2024ReassessingAdvantageIndefinite}.
More strikingly, super-Heisenberg scaling was demonstrated in a continuous-variable
(CV) system using a quantum switch~\citep{zhao2020QuantumMetrologyIndefinite}.
On the other hand, for unitary-channel metrology,
no-advantage results have been established for important higher-order
subclasses: causal-superposition ($\mathsf{CS}$) strategies  do
not outperform parallel strategies~\citep{kurdzialek2023UsingAdaptivenessCausal},
and quantum circuits with quantum control of causal order (QC-QC)
using identical unitary queries can be simulated by sequential circuits~\citep{abbott2024QuantumQueryComplexity}.
These results raise a fundamental question: what are the ultimate
precision limits of ICO quantum metrology?

In this Letter, we address this question for local
frequentist estimation of a single parameter encoded in $N$ uses
of a finite-dimensional quantum channel. In particular, we consider
whether general ICO strategies can exceed Heisenberg scaling, restore
it for channels otherwise confined to the SQL, or improve the leading
QFI coefficient. All asymptotic results refer to
the multi-query limit $N\to\infty$. First, we establish a universal
bound that precludes  QFI scaling of all general ICO processes beyond
the HL. For unitary channels, this bound can be attained by optimal
parallel strategies, showing that their optimal QFI coincides exactly
with that attainable by general ICO strategies. This
extends previous no-advantage results for unitary channels from $\mathsf{CS}$
and QC-QC strategies to the general ICO class, and reveals that nonunitarity
is necessary for ICO metrological advantage. Second, we derive a structurally
refined bound showing that channels restricted to SQL scaling under
parallel strategies remain SQL-limited under general ICO strategies.
It also identifies how the channel structure suppresses the quadratic
contribution and fixes the leading linear coefficient. Furthermore,
an asymptotically tight (AT) bound is  established that determines
the ultimate leading-order precision over the full general ICO  class.
Since parallel strategies asymptotically attain this leading term,
they achieve the ultimate precision permitted by general ICO in both
the HL and SQL regimes. Hence, ICO provides no asymptotic metrological
advantage, and any potential benefit would be confined to finite-query
or subleading-order effects. Finally, for the operationally motivated
QC-QC subclass, we extend the iterative construction for $\mathsf{CS}$
strategies~\citep{kurdzialek2023UsingAdaptivenessCausal} to constrain
finite-query precision, providing a concrete benchmark consistent
with the general-ICO asymptotic result.

\textit{Metrological strategies---}Consider the task of estimating
a parameter $g$ encoded in a quantum channel $\mathcal{E}_{g}\colon L(T^{\mathrm{i}})\to L(T^{\mathrm{o}})$,
where $L(\mathcal{H})$ denotes the set of linear operators on a Hilbert
space $\mathcal{H}$, and $T^{\mathrm{i}}$ and $T^{\mathrm{o}}$
are the input and output spaces of the channel. Via the Choi-Jamiołkowski
isomorphism, a Choi matrix $\mathsf{E}_{g}\in L(T^{\mathrm{i}}\otimes T^{\mathrm{o}})$
represents this channel. For $N$ uses of the channel, the total input
and output spaces are $T^{\mathrm{i}}_{\text{tot}}=\bigotimes^{N}_{i=1}T^{\mathrm{i}}_{i}$
and $T^{\mathrm{o}}_{\text{tot}}=\bigotimes^{N}_{i=1}T^{\mathrm{o}}_{i}$.
A general $N$-party metrological strategy is described by a process
matrix $W\in L(T^{\mathrm{i}}_{\text{tot}}\otimes T^{\mathrm{o}}_{\text{tot}}\otimes F)$,
a positive semidefinite operator connecting the channel's input and
output spaces to a global future space $F$ where the final measurement
occurs. Physically, $W$ encapsulates all possible spatio-temporal
correlations between the channels. As established in Ref.~\citep{liu2023OptimalStrategiesQuantum},
the QFI optimization for any such strategy depends only on the reduced
operator $S=\mathrm{tr}_{F}W$, which acts on the space $L(T^{\mathrm{i}}_{\text{tot}}\otimes T^{\mathrm{o}}_{\text{tot}})$.
We define the set of all such reduced operators $S$ as the strategy
set $\mathsf{S}$.

\begin{figure}[t]
\centering \subfloat{\includegraphics[width=0.8\columnwidth]{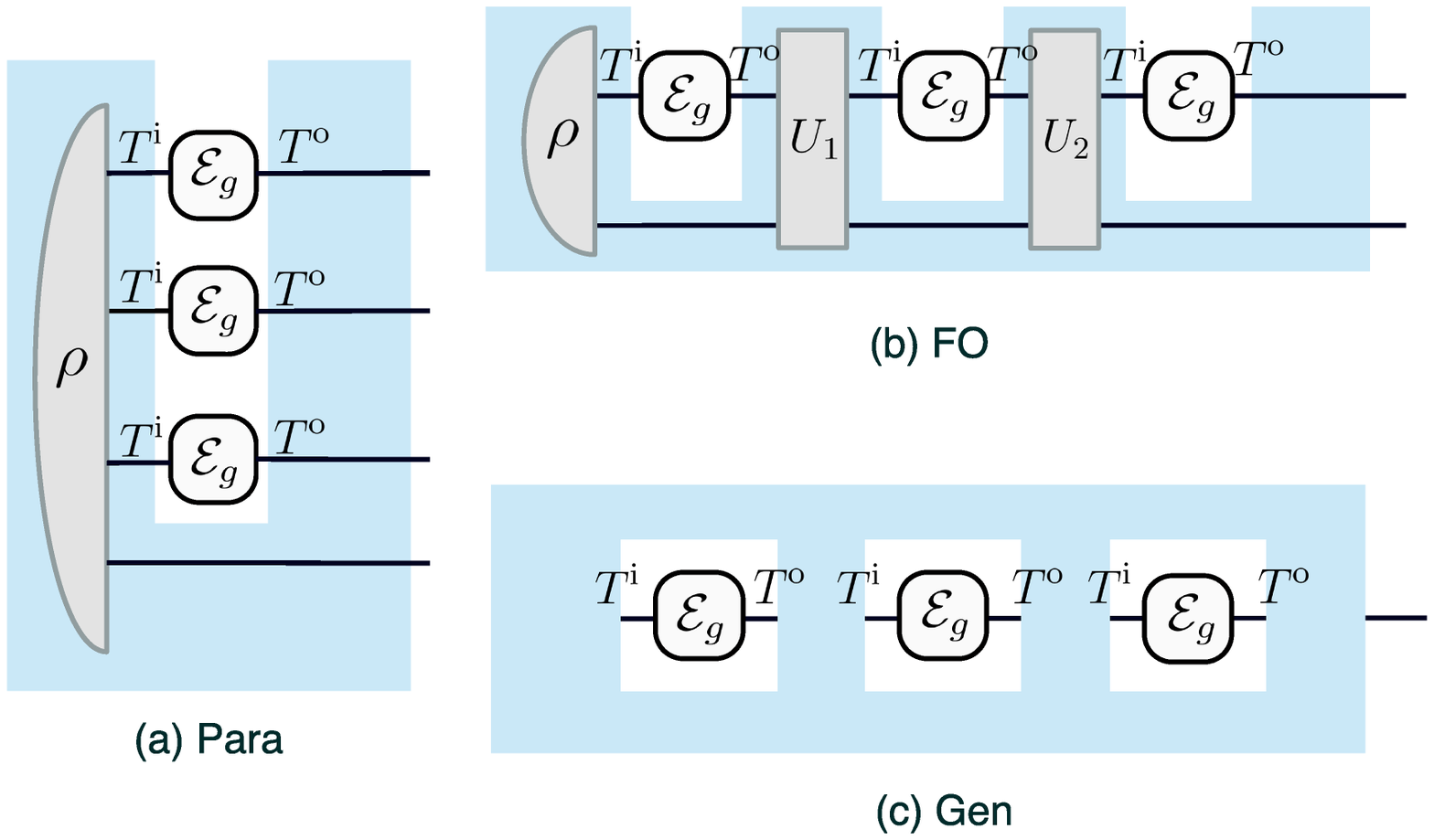} \label{fig:strategies_schematic_sub}}\hspace{0.1\textwidth}
\subfloat{\includegraphics[width=0.65\columnwidth]{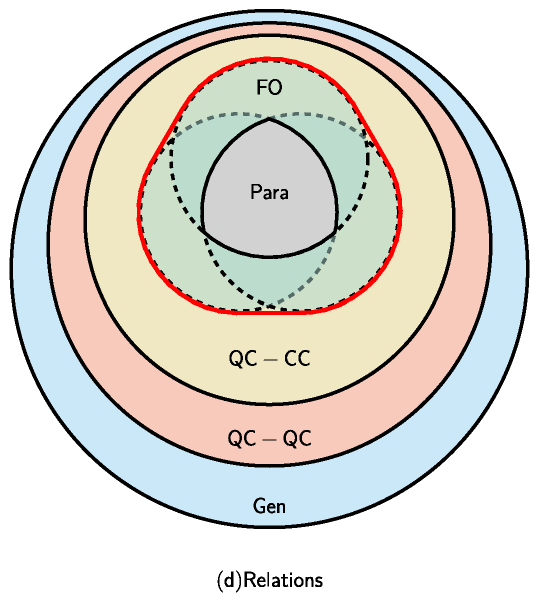}}\caption{Schematic of metrological strategies with different causal structures.
(a) Parallel strategies ($\mathsf{Para}$), where all channels act
simultaneously on an initial state $\rho$. (b) Fixed-order ($\mathsf{FO}$)
strategies with intermediate adaptive operations $U_{i}$ between
channel uses. (c) General ICO strategies ($\mathsf{Gen}$). (d) Hierarchy
of strategy classes.}
\label{fig:strategies_schematic}
\end{figure}

The most general strategies are those allowing ICO, denoted $\mathsf{Gen}$,
as depicted in Fig.~\ref{fig:strategies_schematic}c. The operator
$S$ must satisfy positivity and normalization constraints to avoid
causal paradoxes and guarantee valid probability distributions according
to the generalized Born rule~\citep{oreshkov2012QuantumCorrelationsNo}.
Formally, it must lie in the kernel of the projection superoperator
$\mathcal{Q}_{\mathsf{Gen}}$~\citep{araujo2015WitnessingCausalNonseparability}:
\begin{equation}
S\succeq0,\quad\mathcal{Q}_{\mathsf{Gen}}(S)=0,\quad\mathrm{tr}~S=d^{N},\label{eq:nbodyprocessmatrix}
\end{equation}
where $d$ is the output dimension of a single channel. The explicit
algebraic structure of $\mathcal{Q}_{\mathsf{Gen}}$ is detailed in
Sec.~\ref{sec:ICO-fractal} of the Supplemental Material.

In contrast, the simplest protocols are parallel strategies, denoted
$\mathsf{Para}$ and shown in Fig.~\ref{fig:strategies_schematic}a,
defined by operators of the form $S=\rho\otimes\mathbb{I}_{T^{\mathrm{o}}_{\text{tot}}}$,
where $\rho$ is the input $N$-probe state. This set is characterized
by the projector $\mathcal{Q}_{\mathsf{Para}}=\mathcal{I}-\mathcal{D}_{T^{\mathrm{o}}_{\text{tot}}}$,
where $\mathcal{I}$ is the identity channel and $\mathcal{D}_{T^{\mathrm{o}}_{\text{tot}}}$
is the completely depolarizing channel on $T^{\mathrm{o}}_{\text{tot}}$.
Since $\mathsf{Para}\subset\mathsf{Gen}$, we have $\ker\mathcal{Q}_{\mathsf{Para}}\subseteq\ker\mathcal{Q}_{\mathsf{Gen}}$,
or equivalently, $\mathcal{Q}_{\mathsf{Gen}}(\mathcal{I}-\mathcal{Q}_{\mathsf{Para}})=0$.
This inclusion is strict for $N>1$, while the two sets are identical
for $N=1$. More generally, Fig.~\ref{fig:strategies_schematic}d
shows a finer hierarchy of strategies, implying that any QFI upper
bound derived for $\mathsf{Gen}$ automatically applies to all intermediate
ones, including fixed-order adaptive protocols and quantum circuits
with classical or quantum control of causal order.

The precision for estimating $g$ is quantified by the QFI~\citep{schervish1995TheoryStatistics,braunstein1994StatisticalDistanceGeometry}.
For a strategy $W$, the final state after $N$ channel uses is given
by the link product~\citep{chiribella2009TheoreticalFrameworkQuantum}
$\rho_{g}=W\star\mathsf{E}^{\otimes N}_{g}$. The maximal QFI for
a strategy set $\mathsf{S}$ is found by maximizing over all strategies
$S=\mathrm{tr}_{F}W\in\mathsf{S}$. This can be expressed as a max-min
optimization problem, where the maximization is over $S$ and the
minimization is over all possible purifications of $\rho_{g}$, parameterized
by a Hermitian matrix $h$~\citep{liu2023OptimalStrategiesQuantum}:
\begin{equation}
\mathcal{F}^{\mathsf{S}}_{N}(\mathcal{E}_{g})=\max_{S\in\mathsf{S}}\min_{h\in\mathbb{H}_{r^{N}}}\operatorname{tr}\left(\Omega_{g}(h)S\right),\label{eq:straqfi}
\end{equation}
where $\Omega_{g}(h)=4((\dot{\mathbf{C}}^{0}_{g}-i\mathbf{C}^{0}_{g}h)(\dot{\mathbf{C}}^{0}_{g}-i\mathbf{C}^{0}_{g}h)^{\dagger})^{\top}$
is the ``performance operator''. The matrix $\mathbf{C}^{0}_{g}$
is a matrix derived from an ensemble decomposition of the $N$-channel
Choi matrix, $\mathsf{E}^{\otimes N}_{g}=\sum^{r^{N}}_{j=1}|C_{g,j}\rangle\langle C_{g,j}|=\mathbf{C}^{0}_{g}(\mathbf{C}^{0}_{g})^{\dagger}$,
where the columns of $\mathbf{C}^{0}_{g}$ form a
minimal ensemble decomposition and $r=\operatorname{rank}\mathsf{E}_{g}$
is the single-channel Choi rank. Here $\mathbb{H}_{m}$ denotes the
space of $m\times m$ Hermitian matrices. The dot denotes differentiation
with respect to $g$. For brevity, the subscript $g$ will be omitted
hereafter whenever the context is clear.

\textit{Universal Heisenberg bound---} To determine the ultimate
precision limit of indefinite causal order, we derive a universal
upper bound on the QFI applicable to any quantum channel under any
general ICO strategy ($\mathsf{Gen}$).

\textit{Theorem 1 (Universal HL Bound)---}For any finite-dimensional
quantum channel $\mathcal{E}_{g}$, the maximal QFI achievable under
any indefinite causal order strategy is bounded by the Heisenberg
scaling: 
\begin{equation}
\mathcal{F}^{\mathsf{Gen}}_{N}(\mathcal{E}_{g})\le4N^{2}\min_{\{K_{i}\}}\|\alpha\|,\label{eq:bound1}
\end{equation}
where the minimization is over all Kraus representations $\{K_{i}\}$
of the channel, and $\alpha=\sum_{i}\dot{K}^{\dagger}_{i}\dot{K}_{i}$.
$\|\cdot\|$ denotes the operator norm (the largest singular value).

The proof is based on an iterative decomposition of the $N$-channel
performance operator $\Omega^{(N)}$. Following Ref.~\citep{fujiwara2008FibreBundleManifolds},
we use the specific gauge choice $h^{(N)}=h^{(N-1)}\otimes\mathbb{I}+\mathbb{I}\otimes h^{(1)}$
under which $\Omega^{(N)}$ can be decomposed as 
\begin{equation}
\Omega^{(N)}=\Omega^{(N-1)}\otimes\mathsf{E}^{\top}+\mathsf{E}^{\otimes(N-1)\top}\otimes\Omega^{(1)}+\Omega_{\text{int}},
\end{equation}
where the first term represents the accumulation contribution of $N-1$
channels, the second term represents the local contribution of the
$N$-th channel, and $\Omega_{\text{int}}=\frac{1}{4}(\Lambda^{(N-1)}\otimes\Lambda^{(1)\dagger}+\text{h.c.})$
captures the interference, where h.c. denotes the Hermitian conjugate.
Here, $\Lambda^{(1)}:=4\sum_{i}|\dot{K}^{\dagger}_{i}\rangle\langle K^{\dagger}_{i}|$
is the single-channel interference operator, and $\Lambda^{(N-1)}$
is defined similarly for $N-1$ channels (see Sec.~\ref{sec:appendix_b}
of the Supplemental Material for details). Note that this choice of
$h^{(N)}$ is not generally optimal, but the minimization defining
the QFI allows any admissible gauge to be used to obtain a valid upper
bound.

For any strategy $S\in\mathsf{Gen}$, the QFI is bounded by $\mathcal{F}^{(N)}(S)=\mathrm{tr}(S\Omega^{(N)})$,
which can be accordingly decomposed into three components: an accumulation
term: $\mathsf{H}_{N}(S)=\mathrm{tr}[S(\Omega^{(N-1)}\otimes\mathsf{E}^{\top})]$,
a local term $\mathsf{L}_{N}(S)=\mathrm{tr}[S(\mathsf{E}^{\otimes(N-1)\top}\otimes\Omega^{(1)})]$,
and an interference term $\mathcal{F}_{\text{int}}(S)=\mathrm{tr}(S\Omega_{\text{int}})$.
The key challenge is to bound the interference term. As detailed in
Sec.~\ref{sec:appendix_b} of the Supplemental Material, by applying
a generalized Cauchy-Schwarz inequality, we derive a rigorous bound:
\begin{equation}
|\mathcal{F}_{\text{int}}(S)|\le2\sqrt{\mathsf{H}_{N}(S)\cdot\mathsf{L}_{N}(S)}.
\end{equation}
Substituting this back into the QFI expression yields the single-step
inequality $\mathcal{F}^{(N)}(S)\le(\sqrt{\mathsf{H}_{N}(S)}+\sqrt{\mathsf{L}_{N}(S)})^{2}$.
Finally, we maximize over the set of strategies $\mathsf{Gen}$, and
minimize over channel representations. Because taking the link product
of a valid ICO process matrix with a channel's Choi matrix yields
another valid reduced process matrix, the maximal accumulation and
local QFI contributions are bounded by $\max_{S}\min_{h}\mathsf{H}_{N}(S)\le\mathcal{F}^{\mathsf{Gen}}_{N-1}$
and $\max_{S}\min_{h}\mathsf{L}_{N}(S)\le\mathcal{F}^{\mathsf{Gen}}_{1}$,
respectively. This leads to the recurrence relation $\sqrt{\mathcal{F}^{\mathsf{Gen}}_{N}}\le\sqrt{\mathcal{F}^{\mathsf{Gen}}_{N-1}}+\sqrt{\mathcal{F}^{\mathsf{Gen}}_{1}}$,
which produces the quadratic scaling $\mathcal{F}^{\mathsf{Gen}}_{N}\le N^{2}\mathcal{F}^{\mathsf{Gen}}_{1}=4N^{2}\min_{\{K_{i}\}}\|\alpha\|$.

\textit{Corollary 1 (Unitary Equivalence)---}For a unitary channel
$U_{g}$, the optimal ICO strategy yields the same QFI as the optimal
parallel strategy: 
\begin{equation}
\mathcal{F}^{\mathsf{Gen}}_{N}(U_{g})=\mathcal{F}^{\mathsf{Para}}_{N}(U_{g})=N^{2}(\lambda_{\max}-\lambda_{\min})^{2},
\end{equation}
where $\lambda_{\max}$ and $\lambda_{\min}$ are the maximum and
minimum eigenvalues of the generator $H_{g}=i\dot{U}^{\dagger}_{g}U_{g}$.
Thus, ICO provides no metrological advantage for unitary channels.
This result extends previous no-advantage results
for causal superpositions and QC-QC strategies~\cite{kurdzialek2023UsingAdaptivenessCausal,abbott2024QuantumQueryComplexity}
to the full process-matrix class $\mathsf{Gen}$ optimized here.

\textit{Proof---}For a unitary channel, the optimal parallel strategy
achieves the HL~\citep{pang2014QuantumMetrologyGeneral} with $\mathcal{F}^{\mathsf{Para}}_{N}(U_{g})=N^{2}(\lambda_{\max}-\lambda_{\min})^{2}$.
The minimization in Theorem~1 yields $4\min_{\{K_{i}\}}\|\alpha\|=(\lambda_{\max}-\lambda_{\min})^{2}=\mathcal{F}_{1}(U_{g})$.
This gives $\mathcal{F}^{\mathsf{Gen}}_{N}(U_{g})\le N^{2}\mathcal{F}_{1}(U_{g})=\mathcal{F}^{\mathsf{Para}}_{N}(U_{g})$.
Since $\mathsf{Para}\subset\mathsf{Gen}$, the reverse inequality
follows immediately.\hfill{}$\blacksquare$

Remarkably, Theorem 1 and Corollary 1 are logically equivalent. While
the latter was derived from the former, the converse follows from
the Stinespring dilation theorem~\citep{paulsen2002CompletelyBoundedMaps},
since any ICO strategy on a general channel can be viewed as a restricted
strategy on its unitary purification. Hence, the absence of ICO advantage
for unitary channels implies the universal Heisenberg bound for arbitrary
channels. We provide an independent proof of Corollary
1 in Sec.~\ref{sec:kkt} of the Supplemental Material by using the
Karush-Kuhn-Tucker (KKT) approach \citep{hayashi2024FindingOptimalProbe}
to certify the optimality of the parallel strategy for unitary channels,
from which Theorem 1 can be recovered.

Despite its universality, the HL bound can be loose for some noisy
channels, and may fail to capture the correct asymptotic scaling.
In particular, Pauli channels remain restricted to the SQL even under
optimal ICO strategies, as shown in Sec.~\ref{sec:paulichannel}
of the Supplemental Material using the KKT conditions, although Theorem~1
does not rule out quadratic scaling. This discrepancy raises a key
question: can ICO restore Heisenberg scaling for channels that are
SQL-limited under parallel strategies? To answer this, we derive a
structurally refined bound that captures the combined role of parameter
encoding and noise.

\textit{Theorem 2 (Structurally Refined Bound)---}For any quantum
channel $\mathcal{E}_{g}$, the maximal QFI under any general ICO
strategy ($\mathsf{Gen}$) is bounded by 
\begin{equation}
\begin{split}\mathcal{F}^{\mathsf{Gen}}_{N}(\mathcal{E}_{g})\le\min_{\{K_{i}\}}\bigg[ & 4N\|\alpha\|+N(N-1)\\
 & \times\left(4\|\beta\|^{2}+2d\|\beta\|\|\Lambda^{\perp}\|\right)\bigg],
\end{split}
\label{eq:bound2}
\end{equation}
where $\alpha=\sum_{i}\dot{K}^{\dagger}_{i}\dot{K}_{i}$ and $\beta=\sum_{i}\dot{K}^{\dagger}_{i}K_{i}$.
The term $\Lambda^{\perp}=\Lambda-\Lambda^{\parallel}$ is the orthogonal
component of the single-channel interference operator $\Lambda=4\sum_{i}|\dot{K}^{\dagger}_{i}\rangle\langle K^{\dagger}_{i}|$,
while $\Lambda^{\parallel}=4\beta\otimes\frac{\mathbb{I}_{\mathrm{out}}}{d}$
is the parallel component.

The proof follows from an explicit decomposition of the $N$-channel
performance operator into local contributions scaling as $O(N)$ and
interference contributions scaling as $O(N^{2})$. The key to the
proof lies in analyzing the nonlocal terms via the single-channel
operator $\Lambda$. We find that for any valid two-party ICO process,
the contraction with the purely orthogonal term, $\Lambda^{\perp}\otimes(\Lambda^{\perp})^{\dagger}$,
is identically zero. Consequently, all surviving terms in the $O(N^{2})$
sector of the QFI must involve at least one factor of $\|\beta\|$.
This ensures that if $\beta$ vanishes, the $N^{2}$ scaling vanishes
as well. The detailed derivation is presented in Sec.~\ref{sec:bound2_derivation}
of the Supplemental Material.

Consider channels admitting a Kraus representation such that $\beta=0$
(the Hamiltonian-in-Kraus-span condition~\citep{fujiwara2008FibreBundleManifolds,zhou2021AsymptoticTheoryQuantum}).
This condition is known to restrict parallel strategies to the SQL~\citep{zhou2021AsymptoticTheoryQuantum}.
For such channels, every term in the $O(N^{2})$ part of the refined
bound contains a factor of $\|\beta\|$, making the entire quadratic
term vanish. This implies the asymptotic limit: 
\begin{equation}
\lim_{N\to\infty}{\mathcal{F}^{\mathsf{Gen}}_{N}}/{N}=4\min_{\{K_{i}\}:\beta=0}\|\alpha\|,
\end{equation}
which confirms that ICO cannot attain Heisenberg scaling for channels
otherwise limited to the SQL. In this regime, the refined bound is
asymptotically tight and stronger than the HL bound.

For channels satisfying $\min_{\{K_{i}\}}\|\beta\|\neq0$, a remaining
question is whether general ICO strategies can improve the leading
$N^{2}$ coefficient over the optimal parallel strategy. Theorem~2
does not settle this question because its mixed term $2d\|\beta\|\|\Lambda^{\perp}\|$
is obtained by bounding channel-use pairs separately. The following
bound instead treats all $N$ uses jointly.

\textit{Theorem 3 (Asymptotically Tight Bound)---}For
any finite-dimensional quantum channel $\mathcal{E}_{g}$, the maximal
QFI over general ICO strategies satisfies 
\begin{equation}
\mathcal{F}^{\mathsf{Gen}}_{N}(\mathcal{E}_{g})\le4\min_{\{K_{i}\}}\left[N\|\beta\|+\sqrt{N\|\alpha-\beta\beta^{\dagger}\|}\right]^{2},\label{eq:bound3}
\end{equation}
where $\alpha-\beta\beta^{\dagger}\succeq0$ for every Kraus representation
in the minimization.

To outline the proof, we collect the Kraus operators
into a Stinespring isometry $V:=\sum_{i}|i\rangle_{E}\otimes K_{i}$
with the tangent vector $\dot{V}:=\sum_{i}|i\rangle_{E}\otimes\dot{K}_{i}$,
and decompose it as $\dot{V}=V\beta^{\dagger}+R$, where $R:=\sum_{i}|i\rangle_{E}\otimes R_{i}$
and $R_{i}:=\dot{K}_{i}-K_{i}\beta^{\dagger}$. The term $V\beta^{\dagger}$
can accumulate coherently over the $N$ channel uses and contributes
at most $N\|\beta\|$ to Eq.~\eqref{eq:bound3}. By differentiating
the completeness relation $V^{\dagger}V=\mathbb{I}_{A}$, the component
$R$ satisfies $R^{\dagger}V=0$. This implies that $4\sum_{i}|R^{\dagger}_{i}\rangle\langle K^{\dagger}_{i}|$
has zero partial trace over the channel output. As in the proof of
Theorem 2, the tensor product of two operators with vanishing partial
trace over the output space has zero inner product with any valid
two-party process matrix. Hence, the terms carrying a factor of $N$
cancel out. The remaining residual contribution is bounded by $\sqrt{N\|\alpha-\beta\beta^{\dagger}\|}$
in Eq.~\eqref{eq:bound3}. A complete proof is given in Sec.~\ref{sec:ultimate_bound}
of the Supplemental Material.

If the channel admits a Kraus representation with
$\beta=0$, Eq.~\eqref{eq:bound3} gives $\mathcal{F}^{\mathsf{Gen}}_{N}\le4N\min_{\{K_{i}\}:\beta=0}\|\alpha\|$
for every $N$, recovering the linear finite-query bound of Theorem~2.
Together with the known asymptotic attainability by parallel strategies~\citep{zhou2021AsymptoticTheoryQuantum},
this also fixes the same asymptotic linear coefficient. In the complementary
regime $\min_{\{K_{i}\}}\|\beta\|>0$, parallel strategies attain
the quadratic coefficient appearing in Eq. \eqref{eq:bound3}, and
the inclusion relation $\mathsf{Para}\subset\mathsf{Gen}$ therefore
gives 
\begin{equation}
\lim_{N\to\infty}\frac{\mathcal{F}^{\mathsf{Gen}}_{N}}{N^{2}}=\lim_{N\to\infty}\frac{\mathcal{F}^{\mathsf{Para}}_{N}}{N^{2}}=4\min_{\{K_{i}\}}\|\beta\|^{2}.\label{eq:general_asymptotic_equivalence}
\end{equation}
Consequently, in both the SQL and HL regimes, $\mathcal{F}^{\mathsf{Gen}}_{N}/\mathcal{F}^{\mathsf{Para}}_{N}\to1$,
whenever the common leading coefficient is nonzero. General ICO strategies
can therefore improve neither the asymptotic scaling nor its leading
coefficient, although finite-query or subleading enhancements may
remain. At finite $N$, the three bounds are complementary rather
than uniformly ordered: Theorem 1 provides the simplest universal
Heisenberg-scaling envelope, Theorem 2 reveals a noise-sensitive interference
structure underlying the stability of SQL scaling, and Theorem 3 determines
the leading asymptotic coefficient in both the SQL and HL regimes.

For a unitary channel, the second term in Eq.~\eqref{eq:bound3}
vanishes for a single-Kraus representation, so Theorem~3 is reduced
exactly to the universal bound of Theorem~1, which is attained by
an optimal parallel strategy.

\textit{\emph{Although QC-QC strategies are already covered by Theorem
3 as a subclass of $\mathsf{Gen}$, their generalized circuit structure
permits a recursive finite-query analysis.}}\emph{ }By extending the
method of Ref.~\citep{kurdzialek2023UsingAdaptivenessCausal}, we
prove that QC-QC strategies satisfy the same finite-query iterative
bound as $\mathsf{CS}$ strategies and reproduce the leading asymptotic
behavior of Theorem~3. Moreover, since $\mathsf{Para}\subseteq\mathsf{CS}\subseteq\mathsf{QC\text{-}QC}\subseteq\mathsf{Gen}$,
the asymptotic equivalence between parallel and general-ICO strategies
established by Theorem 3 implies that all four classes share the same
leading QFI coefficient. Therefore, when $N\rightarrow\infty$, $\mathcal{F}^{\mathsf{QC\text{-}QC}}_{N}/\mathcal{F}^{\mathsf{Para}}_{N}\to1$.
The complete derivation is given in Sec.~\ref{sec:qcqc} of the Supplemental
Material.
\begin{figure}[t]
\centering \includegraphics[width=1\columnwidth]{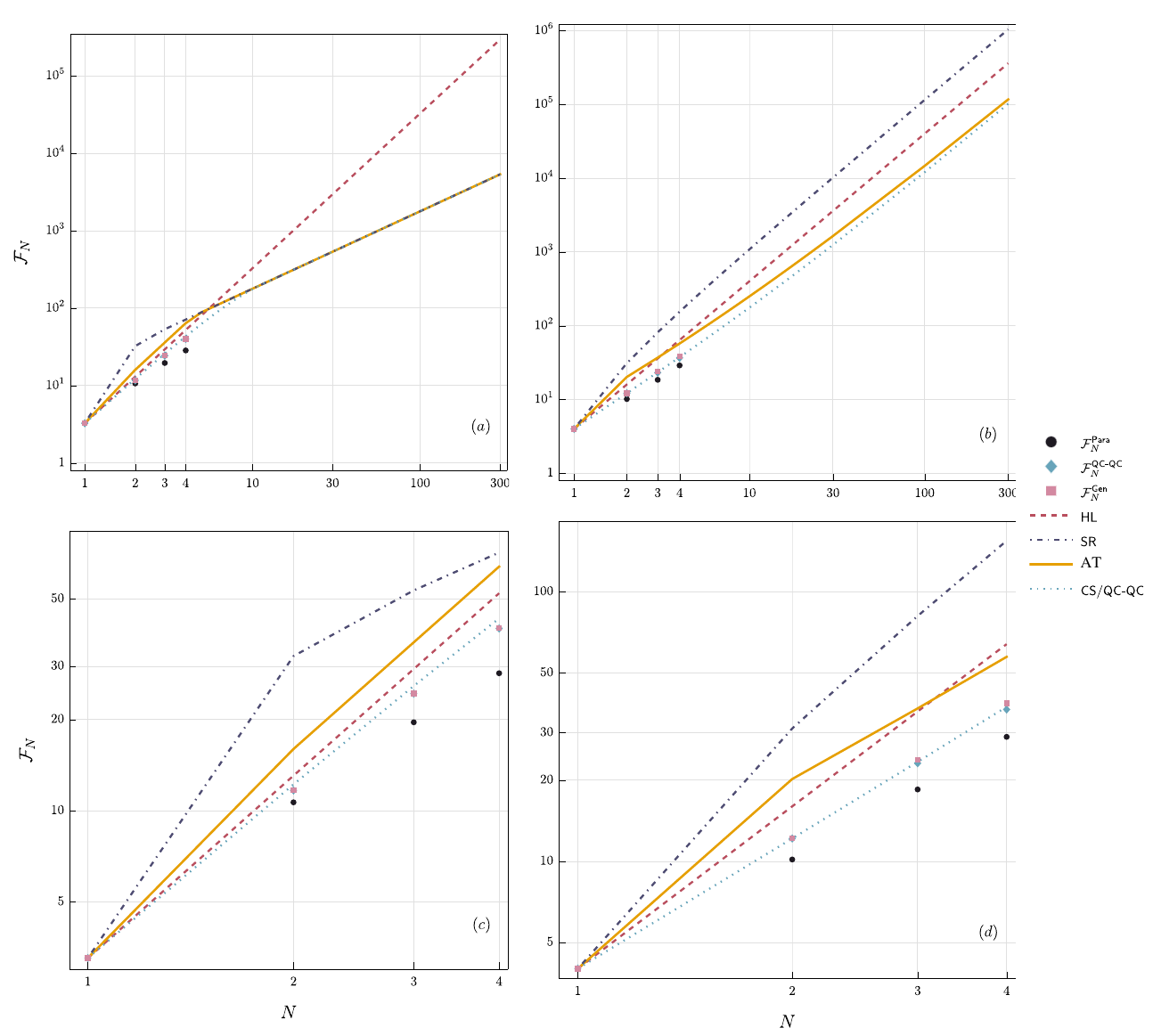}
\caption{Numerical comparison of the optimized QFI $\mathcal{F}_{N}$ versus
the number of channel uses $N$ for parallel, QC--QC, and general-ICO
strategies, together with the HL, SR, AT, and CS/QC--QC iterative
bounds. Panels (a) and (c) consider phase estimation under a generalized
amplitude-damping (GAD)  channel described by Kraus operators $E_{0}=\cos\theta\left(|0\rangle\langle0|+\sqrt{1-p}|1\rangle\langle1|\right)$,
$E_{1}=\cos\theta\sqrt{p}|0\rangle\langle1|$, $E_{2}=\sin\theta\left(\sqrt{1-p}|0\rangle\langle0|+|1\rangle\langle1|\right)$,
and $E_{3}=\sin\theta\sqrt{p}|1\rangle\langle0|$ with $p=0.31$ and
$\theta=\pi/4$, and the parameter is encoded by $U_{g}=\exp(-ig\sigma_{3})$
with  $g=1/2$ before the noise. This channel satisfies the Hamiltonian-in-Kraus-span
condition and is therefore SQL-limited. Panels (b) and (d) consider
a perpendicular-dephasing channel followed by the same unitary encoding
$U_{g}$, with parameter-dependent Kraus operators $K_{0}(g)=\sqrt{p}U_{g}$
and $K_{1}(g)=\sqrt{1-p}U_{g}\sigma_{1}$ where $p=0.76$. The orthogonality
between the signal and noise axes yields $\min_{\{K_{i}\}}\|\beta\|\protect\neq0$,
allowing HL scaling. The CS/QC--QC iterative bound applies to the
QC--QC strategies only.}
\label{fig:numerical_examples}
\end{figure}

\textit{Numerical Examples---}We numerically optimize the QFI attainable
by parallel, QC--QC, and general ICO strategies using the semidefinite-programming
methods of Refs.~\citep{liu2023OptimalStrategiesQuantum,hayashi2023TightCramerRaoType,
hayashi2024FindingOptimalProbe,mothe2024ReassessingAdvantageIndefinite}. Fig.~\ref{fig:numerical_examples} compares the results with the
universal HL, structurally refined (SR), AT, and CS/QC--QC iterative
bounds for two typical noisy channels: generalized amplitude damping
in the SQL regime and perpendicular dephasing in the HL regime. In
both examples, the general ICO QFI satisfies all three general bounds,
while the QC--QC QFI satisfies the corresponding iterative bound
within numerical precision. For perpendicular dephasing, the general
ICO QFI exceeds the CS/QC--QC iterative bound at $N=3$ and $4$,
illustrating that the latter applies specifically to the CS and QC--QC
subclasses rather than to arbitrary ICO processes.

\textit{Conclusion---}We have established universal asymptotic precision
limits for quantum channel estimation beyond fixed causal order. General
ICO strategies in finite-dimensional systems remain bounded by Heisenberg
scaling, and for unitary channels their optimal QFI coincides with
that of parallel strategies. For noisy channels, the optimal parallel
and general ICO strategies share the leading asymptotic order of the
QFI with the same coefficient in both the SQL and HL regimes. Thus,
any ICO advantage is confined to finite-query or subleading effects,
and may also arise beyond the scope of present problem, including
Bayesian estimation~\cite{bavaresco2024DesigningOptimalProtocolsa},
multiparameter sensing~\cite{hayashi2023TightCramerRaoType}, non-Markovian
metrology~\cite{altherr2021QuantumMetrologyNonMarkovian}, and other
information-processing tasks such as channel discrimination and exclusion~\cite{fang2025GeneralizedQuantumChernoff},
etc.

\textit{Acknowledgments---} The authors acknowledge Runjian Huang
and Xiyuan Xie for valuable discussions and suggestions. This work
is supported by the National Natural Science Foundation of China (Grant
No. 12075323), the Natural Science Foundation of Guangdong Province
of China (Grant No. 2025A1515011440), and the Innovation Program for
Quantum Science and Technology (Grant No. 2021ZD0300702).

\sloppy
\emergencystretch=3em
\Urlmuskip=0mu plus 1mu
\bibliography{projectICOmetrologyref}
\fussy
\emergencystretch=0pt
\Urlmuskip=0mu

\onecolumngrid \clearpage{}

\setcounter{equation}{0} \setcounter{page}{1} \setcounter{figure}{0}
\setcounter{table}{0} \makeatletter 
\global\long\def\theequation{S\arabic{equation}}%
\global\long\def\thefigure{S\arabic{figure}}%
\global\long\def\thetable{S\arabic{table}}%
\global\long\def\theHequation{S\arabic{equation}}%
\global\long\def\theHfigure{S\arabic{figure}}%
\global\long\def\theHtable{S\arabic{table}}%
\global\long\def\theHsection{S\arabic{section}}%
\setcounter{secnumdepth}{2}
\begin{center}
\vspace*{0.5em}
{\Large\bfseries Supplemental Material for\\[0.3em]
``Fundamental Limits of Quantum Metrology Beyond Fixed Causal Order''\par}
\vspace{1em}
{\large Wenjie Wei$^{1}$\orcidlink{0009-0001-2435-0852},
Yutong Li$^{1}$\orcidlink{0009-0003-9365-6913}, and
Shengshi Pang$^{1,2}$\orcidlink{0000-0002-6351-539X}\par}
\vspace{0.5em}
{\small $^{1}$School of Physics, Sun Yat-sen University, Guangzhou, Guangdong 510275, China\par
$^{2}$Hefei National Laboratory, University of Science and Technology of China, Hefei 230088, China\par}
\vspace{0.4em}
{\small\href{mailto:pangshsh@mail.sysu.edu.cn}{\texttt{pangshsh@mail.sysu.edu.cn}}\par}
\vspace{0.4em}
{\small\today\par}
\end{center}
\vspace{1.2em}
\onecolumngrid

This Supplemental Material provides the technical foundations and
detailed proofs underlying the results presented in the main text.
We first characterize the algebraic structure of general indefinite-causal-order
(ICO) strategies and reveal an unexpected fractal geometry in the
corresponding process space. We then derive the universal Heisenberg-limit
(HL) bound and the structurally refined bound through complementary
approaches based on iterative inequalities, convex optimization, and
Karush-Kuhn-Tucker (KKT) conditions. We next prove the asymptotically
tight (AT) bound for general ICO strategies, which fixes the same
asymptotic leading quantum Fisher information (QFI) coefficient as
the optimal parallel strategy in both the standard quantum limit (SQL)
and Heisenberg-scaling regimes. We also analyze Pauli channels and
derive an iterative finite-query bound for quantum circuits with quantum
control of causal order (QC-QC). Detailed derivations, auxiliary lemmas,
and numerical methods are provided throughout.

\section{Algebraic Structure of $\mathcal{Q}_{\mathsf{Gen}}$ and Fractal
Patterns in ICO}

\label{sec:ICO-fractal}

As discussed in the main text, a physically valid indefinite causal
order (ICO) process must not allow for causal paradoxes and must yield
normalized probabilities for any local experimenter. Mathematically,
this dictates that the process matrix $S$ must lie in the kernel
of a projection superoperator, $\mathcal{Q}_{\mathsf{Gen}}$. In this
section, we explicitly define this superoperator and reveal the striking
fractal geometry of its image space. For $N$ uses of a quantum channel,
the projection superoperator is explicitly constructed as: 
\begin{equation}
\mathcal{Q}_{\mathsf{Gen}}=\bigotimes^{N}_{i=1}(\mathcal{I}-\mathcal{D}_{T^{\mathrm{o}}_{i}}+\mathcal{D}_{T^{\mathrm{i}}_{i}T^{\mathrm{o}}_{i}})-\mathcal{D}_{T^{\mathrm{i}}_{\mathrm{tot}}T^{\mathrm{o}}_{\mathrm{tot}}},\label{eq:qico_appendix}
\end{equation}
where $\mathcal{I}$ is the identity superoperator, and $\mathcal{D}_{X}(\cdot)=\mathrm{tr}_{X}(\cdot)\otimes\mathbb{I}_{X}/\dim(X)$
denotes the completely depolarizing channel on subsystem $X$. The
characterization of the image space, $\mathrm{im}\mathcal{Q}_{\mathsf{Gen}}$,
is pivotal to the theory of ICO. Operators within
this image space correspond to terms projected out by the validity
constraints, which do not define valid process matrices and would
disrupt the normalization of measurement probabilities. Conversely,
elements in its orthogonal complement, the kernel $\mathrm{ker}\mathcal{Q}_{\mathsf{Gen}}$,
preserve physical consistency~\citep{oreshkov2012QuantumCorrelationsNo}.
We can characterize this space $\mathrm{im}\mathcal{Q}_{\mathsf{Gen}}$
as follows. Let $\mathcal{W}^{0}=\mathbb{C}\mathbb{I}\subset L(\mathcal{H})$
be the space spanned by the single-system identity operator, and let
$\mathcal{W}^{1}=(\mathcal{W}^{0})^{\perp}$ be the space of single-system
traceless operators. The total operator space can then be decomposed
as 
\begin{equation}
\begin{aligned} & L(T^{\mathrm{i}}_{\text{tot}}\otimes T^{\mathrm{o}}_{\text{tot}})=\left(\mathcal{W}^{0}\oplus\mathcal{W}^{1}\right)^{\otimes2N}\\
={} & \bigoplus_{\mathbf{i},\mathbf{j}\in(0,1)^{N}}\mathcal{W}^{i_{1}}_{T^{\mathrm{i}}_{1}}\otimes\cdots\otimes\mathcal{W}^{i_{N}}_{T^{\mathrm{i}}_{N}}\otimes\mathcal{W}^{j_{1}}_{T^{\mathrm{o}}_{1}}\otimes\cdots\otimes\mathcal{W}^{j_{N}}_{T^{\mathrm{o}}_{N}}\\
={} & \bigoplus_{\mathbf{i},\mathbf{j}}\mathcal{W}^{\mathbf{i}}_{{T^{\mathrm{i}}_{\text{tot}}}}\otimes\mathcal{W}^{\mathbf{j}}_{{T^{\mathrm{o}}_{\text{tot}}}},
\end{aligned}
\label{eq:combbasis}
\end{equation}
where $\mathbf{i}\equiv(i_{1},\ldots,i_{N})$ and $\mathbf{j}\equiv(j_{1},\ldots,j_{N})$
are binary vectors, $\mathcal{W}^{\mathbf{i}}=\mathcal{W}^{i_{1}}\otimes\cdots\otimes\mathcal{W}^{i_{N}}$,
and $\mathcal{W}^{\mathbf{j}}=\mathcal{W}^{j_{1}}\otimes\cdots\otimes\mathcal{W}^{j_{N}}$.
The subscripts denote the subsystems. As proven in the Supplemental
Material~\ref{sec:proofofcharaoficq}, the image space of $\mathcal{Q}_{\mathsf{Gen}}$
is 
\begin{equation}
\begin{aligned}\mathrm{im}\mathcal{Q}_{\mathsf{Gen}}=\bigoplus_{\begin{subarray}{c}
(\mathbf{i},\mathbf{j})\neq(0,0)\\
(\mathbf{i}\oplus\mathbf{j})_{(2)}=\mathbf{j}_{(2)}-\mathbf{i}_{(2)}
\end{subarray}}\mathcal{W}^{\mathbf{i}}_{{T^{\mathrm{i}}_{\text{tot}}}}\otimes\mathcal{W}^{\mathbf{j}}_{{T^{\mathrm{o}}_{\text{tot}}}}.\end{aligned}
\label{eq:charaoficq}
\end{equation}
Here, $\mathbf{i}_{(2)}$ is the integer representation of the binary
vector $\mathbf{i}$, with the leftmost bit being the most significant.
Since bitwise exclusive OR (XOR) is non-negative, the summation constraint
implies $\mathbf{i}_{(2)}\leqslant\mathbf{j}_{(2)}$. The equality
constraint in Eq.~\eqref{eq:charaoficq} is equivalent to the concise
condition 
\begin{equation}
i_{k}\leq j_{k},\quad k=1,2,\cdots,N.
\end{equation}
{} This condition has a simple interpretation. Apart
from the trivial identity component, the terms selected by $\operatorname{im}\mathcal{Q}_{\mathsf{Gen}}$
are those in which the output side is nontrivial, either together
with a nontrivial input side or with a trivial input side. Such terms
are not compatible with local quantum mechanics as valid process-matrix
components, since they would make the normalization depend on the
local deterministic operation inserted into the process. Since the
condition $i_{k}\leq j_{k}$ is a purely local bitwise rule, adding
one more channel use simply adds one new binary coordinate. This splits
the previous pattern into four large blocks: three inherit the old
pattern, while the block corresponding to $(i_{k},j_{k})=(1,0)$ is
removed. Repeating the same rule at every scale naturally produces
the Sierpiński-type self-similar structure.

\begin{figure}[htbp]
\centering \includegraphics[height=0.68\textheight]{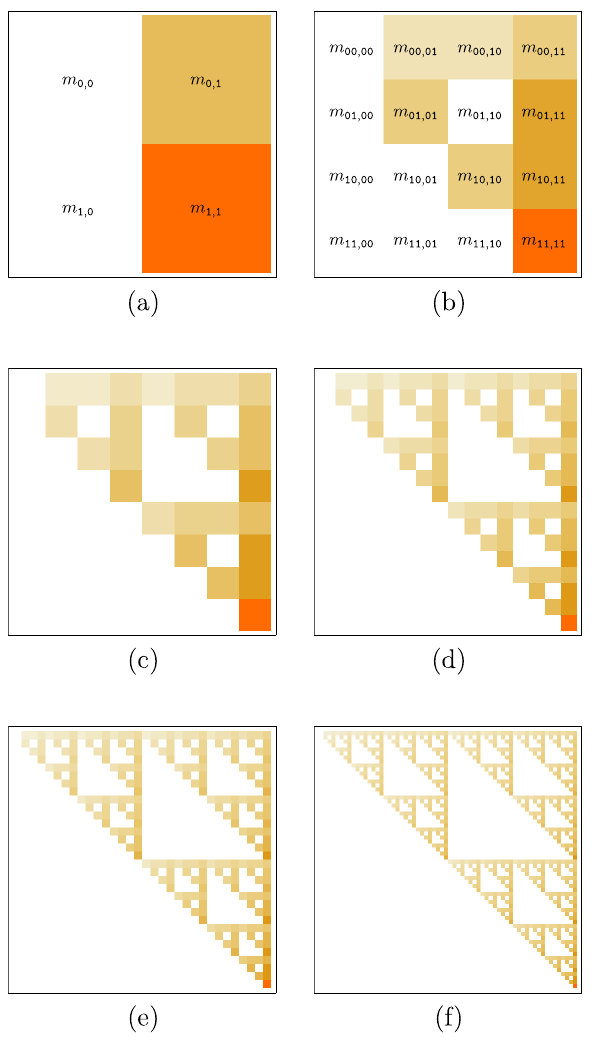} \caption{The Sierpinski triangle in the image space of the ICO projection operator.
Panels (a)-{}-(f) plot the matrix defined in Eq.~\eqref{eq:mij}
for $N=1$ to $N=6$. Each pixel corresponds to a subspace. Blank
pixels (zero elements) represent terms allowed in ICO process matrices,
i.e., in $\ker\mathcal{Q}_{\mathsf{Gen}}$, while colored pixels represent
forbidden terms belonging to the image space $\operatorname{im}\mathcal{Q}_{\mathsf{Gen}}$.
Darker colors indicate higher subspace dimensions.}
\label{fig:ico_mij_matrix}
\end{figure}

This fractal pattern is visualized by defining a matrix $M$. Its
elements are 
\begin{equation}
m_{\mathbf{i},\mathbf{j}}=\begin{cases}
0, & (\mathbf{i},\mathbf{j})=(\mathbf{0},\mathbf{0}),\\
\delta_{(\mathbf{i}\oplus\mathbf{j})_{(2)},\mathbf{j}_{(2)}-\mathbf{i}_{(2)}}\mathrm{dim}\left(\mathcal{W}^{\mathbf{i}}_{{T^{\mathrm{i}}_{\text{tot}}}}\otimes\mathcal{W}^{\mathbf{j}}_{{T^{\mathrm{o}}_{\text{tot}}}}\right), & \text{otherwise},
\end{cases}\label{eq:mij}
\end{equation}
where the dimension of the subspace is $\mathrm{dim}(\mathcal{W}^{\mathbf{i}}_{{T^{\mathrm{i}}_{\text{tot}}}}\otimes\mathcal{W}^{\mathbf{j}}_{{T^{\mathrm{o}}_{\text{tot}}}})=(d^{2}-1)^{\mathrm{wt}(\mathbf{i})+\mathrm{wt}(\mathbf{j})}$,
with $\mathrm{wt}(\mathbf{i})$ being the Hamming weight of $\mathbf{i}$.
Fig.~\ref{fig:ico_mij_matrix} plots this matrix, showing that the
fractional area of the nonzero (colored) region diminishes as $N$
increases, vanishing as $N\to\infty$. This holds even when weighted
by the subspace dimensions. The total dimension of the image space
$\mathrm{im}\mathcal{Q}_{\mathsf{Gen}}$ is obtained by summing over
Eq.~\eqref{eq:mij}: 
\begin{equation}
\sum_{\mathbf{i},\mathbf{j}}m_{\mathbf{i},\mathbf{j}}=\left(d^{4}-d^{2}+1\right)^{N}-1.
\end{equation}
Given that the dimension of the total space $L(T^{\mathrm{i}}_{\text{tot}}\otimes T^{\mathrm{o}}_{\text{tot}})$
is $d^{4N}$, the dimensional fraction of $\mathrm{im}\mathcal{Q}_{\mathsf{Gen}}$
approaches zero as $N\to\infty$ for any $d$. This means that almost
all basis terms in the decomposition of Eq.~\eqref{eq:combbasis}
are permissible in an ICO process matrix. This behavior is in stark
contrast to that of quantum combs, for which almost all terms become
forbidden in the large-$N$ limit.

\subsection{Proof for Characterization of Image Space of $\mathcal{Q}_{\mathsf{Gen}}$
{[}Eq.~\eqref{eq:charaoficq}{]}}

\label{sec:proofofcharaoficq}

\textit{Proof.}--- Let $A^{i_{k}}\in\mathcal{W}^{i_{k}}$ with $i_{k}\in\{0,1\}$,
and similarly for $j_{k}$. From Eq.~\eqref{eq:qico_appendix} and
the definition of $\mathcal{D}_{X}$, the action of $\mathcal{Q}_{\mathsf{Gen}}$
on a basis element factorizes over the $N$ channel slots: 
\begin{equation}
\mathcal{Q}_{\mathsf{Gen}}\bigl(A^{i_{1}}\otimes\cdots\otimes A^{j_{N}}\bigr)=\left[\prod^{N}_{k=1}\bigl(1-\delta_{j_{k}0}+\delta_{i_{k}0}\delta_{j_{k}0}\bigr)-\delta_{\mathbf{i},\mathbf{0}}\delta_{\mathbf{j},\mathbf{0}}\right]\bigl(A^{i_{1}}\otimes\cdots\otimes A^{j_{N}}\bigr).
\end{equation}
Hence a basis subspace $\mathcal{W}^{\mathbf{i}}\otimes\mathcal{W}^{\mathbf{j}}$
lies in $\mathrm{im}\,\mathcal{Q}_{\mathsf{Gen}}$ iff the coefficient
above equals $1$. Thus the all-identity component
$(\mathbf{i},\mathbf{j})=(\mathbf{0},\mathbf{0})$ is removed by the
global depolarizing term, while all other components are selected
exactly when the product equals $1$. Inspecting the $k$-th factor:
\begin{itemize}
\item if $j_{k}=1$: the factor is $1$ regardless of $i_{k}$.
\item if $(i_{k},j_{k})=(0,0)$: the factor is $1-1+1=1$.
\item if $(i_{k},j_{k})=(1,0)$: the factor is $1-1+0=0$.
\end{itemize}
Therefore the product equals $1$ iff for every $k$ we have $i_{k}\leq j_{k}$,
i.e., the pair $(i_{k},j_{k})$ never equals $(1,0)$. Now note that
for binary digits $i_{k},j_{k}\in\{0,1\}$, the relation $i_{k}\leq j_{k}$
is equivalent to $i_{k}\oplus j_{k}=j_{k}-i_{k}$. Consequently, $i_{k}\leq j_{k}\;\;\forall k$
is equivalent to 
\begin{equation}
\sum^{N-1}_{k=0}2^{k}(i_{N-k}\oplus j_{N-k})=\sum^{N-1}_{k=0}2^{k}(j_{N-k}-i_{N-k}),
\end{equation}
which is exactly $(\mathbf{i}\oplus\mathbf{j})_{(2)}=\mathbf{j}_{(2)}-\mathbf{i}_{(2)}$.
Thus $\mathcal{W}^{\mathbf{i}}\otimes\mathcal{W}^{\mathbf{j}}\subset\mathrm{im}\,\mathcal{Q}_{\mathsf{Gen}}$
precisely when $(\mathbf{i},\mathbf{j})\neq(0,0)$ and $(\mathbf{i}\oplus\mathbf{j})_{(2)}=\mathbf{j}_{(2)}-\mathbf{i}_{(2)}$,
proving Eq.~\eqref{eq:charaoficq}. \hfill{}$\blacksquare$

\section{Proof of Theorem 1 (Universal HL Bound)}

\label{sec:appendix_b}

In this section, we provide a detailed proof of the universal HL bound.
This approach leverages a universal single-step iterative inequality
for the QFI, which is derived from the iterative structure of the
$N$-channel performance operator and a generalized operator Cauchy--Schwarz
inequality. The total $N$-channel performance operator $\Omega^{(N)}$
can be expressed iteratively in terms of the $(N-1)$-channel and
single-channel operators. 
\begin{equation}
\begin{aligned}\Omega^{(N)} & =\Omega^{(N-1)}\otimes\mathsf{E}^{\top}+\mathsf{E}^{\otimes(N-1)\top}\otimes\Omega^{(1)}\\
 & +\frac{1}{4}\left(\Lambda^{(N-1)}\otimes\Lambda^{(1)\dagger}+\text{h.c.}\right),
\end{aligned}
\label{eq:apd_omega_iter}
\end{equation}
where we have omitted the explicit dependence on the parameterization
$h$ for brevity. By taking the trace of this operator with a strategy
$S$, we obtain an iterative formula for the QFI: 
\begin{equation}
\mathcal{F}^{(N)}(S)=\mathsf{H}_{N}(S)+\mathsf{L}_{N}(S)+\mathcal{F}_{\text{int}}(S),\label{eq:apd_qfi_iter_full}
\end{equation}
where we have defined the QFI $\mathcal{F}^{(N)}(S)=\mathrm{tr}(S\Omega^{(N)})$,
the accumulation term $\mathsf{H}_{N}(S)=\mathrm{tr}\left[(\Omega^{(N-1)}\otimes\mathsf{E}^{\top})S\right]$,
the local term $\mathsf{L}_{N}(S)=\mathrm{tr}\left[(\mathsf{E}^{\otimes(N-1)\top}\otimes\Omega^{(1)})S\right]$,
and the interference term $\mathcal{F}_{\text{int}}(S)=\frac{1}{4}\mathrm{tr}\left[\left(\Lambda^{(N-1)}\otimes\Lambda^{(1)\dagger}+\text{h.c.}\right)S\right]$.
The core of this proof lies in bounding the interference term. For
any semidefinite operator $S\succeq0$, this term is constrained by
the operator Cauchy-Schwarz inequality in the Hilbert-Schmidt inner
product space. Specifically, we can prove the following inequality:
\begin{equation}
|\mathcal{F}_{\text{int}}(S)|\le2\sqrt{\mathsf{H}_{N}(S)\cdot\mathsf{L}_{N}(S)}.\label{eq:apd_int_bound}
\end{equation}
To prove this, we define two auxiliary operators 
\begin{equation}
\mathsf{A}:=\left(2\mathbf{C}^{(N-1)\dagger}\otimes(\dot{\mathbf{C}}-i\mathbf{C}h)^{(1)\dagger}\right)^{\top}\sqrt{S},
\end{equation}
and 
\begin{equation}
\mathsf{B}:=\left(2(\dot{\mathbf{C}}-i\mathbf{C}h)^{(N-1)\dagger}\otimes\mathbf{C}^{(1)\dagger}\right)^{\top}\sqrt{S}.
\end{equation}
The standard Cauchy-Schwarz inequality states $|\mathrm{tr}(\mathsf{A}^{\dagger}\mathsf{B})|^{2}\le\mathrm{tr}(\mathsf{A}^{\dagger}\mathsf{A})\cdot\mathrm{tr}(\mathsf{B}^{\dagger}\mathsf{B})$.
By using the cyclic property of the trace, one can show that $\mathrm{tr}(\mathsf{A}^{\dagger}\mathsf{B})=\mathrm{tr}[S(\Lambda^{(N-1)}\otimes\Lambda^{(1)\dagger})]/4$,
$\mathrm{tr}(\mathsf{A}^{\dagger}\mathsf{A})=\mathrm{tr}[S(\mathsf{E}^{\otimes(N-1)\top}\otimes\Omega^{(1)})]$,
and $\mathrm{tr}(\mathsf{B}^{\dagger}\mathsf{B})=\mathrm{tr}[S(\Omega^{(N-1)}\otimes\mathsf{E}^{\top})]$.
This directly leads to the inequality $|\mathrm{tr}[S(\Lambda^{(N-1)}\otimes\Lambda^{(1)\dagger})]|^{2}\le16\mathsf{H}_{N}(S)\cdot\mathsf{L}_{N}(S)$.
Substituting this bound into Eq.~\eqref{eq:apd_qfi_iter_full}, we
arrive at a universal single-step inequality for the QFI: 
\begin{equation}
\mathcal{F}^{(N)}(S)\le\left(\sqrt{\mathsf{H}_{N}(S)}+\sqrt{\mathsf{L}_{N}(S)}\right)^{2}.\label{eq:apd_cs_ineq}
\end{equation}
To find the bound on the maximal achievable QFI, $\mathcal{F}^{\mathsf{Gen}}_{N}$,
we take the maximum over all valid strategies $S\in\mathsf{Gen}$
and the minimum over all parameterizations $h$ on both sides of Eq.~\eqref{eq:apd_cs_ineq}.
Since the function $(x,y)\mapsto(\sqrt{x}+\sqrt{y})^{2}$ is monotonically
increasing, we can bound the maximum of the sum by the sum of the
maxima: 
\begin{equation}
\begin{split}\mathcal{F}^{\mathsf{Gen}}_{N} & =\max_{S\in\mathsf{Gen}}\min_{h^{(N)}}\mathcal{F}^{(N)}(S)\\
 & \le\left(\sqrt{\max_{S\in\mathsf{Gen}}\min_{h^{(N-1)}}\mathsf{H}_{N}(S)}+\sqrt{\max_{S\in\mathsf{Gen}}\min_{h^{(1)}}\mathsf{L}_{N}(S)}\right)^{2}.
\end{split}
\label{eq:apd_max_ineq}
\end{equation}
The crucial step is to bound the two maxima on the right-hand side.
This is where the physical constraint of a valid causal structure
becomes essential. For any valid strategy $S\in\mathsf{Gen}$, the
reduced processes corresponding to the accumulation and local terms
must also be valid ICO processes. Consequently, their achievable QFI
cannot exceed the optimal QFI for the corresponding number of channel
uses. This leads to the bounds: 
\begin{align}
\max_{S\in\mathsf{Gen}}\min_{h}\mathsf{H}_{N}(S) & \le\mathcal{F}^{\mathsf{Gen}}_{N-1},\\
\max_{S\in\mathsf{Gen}}\min_{h}\mathsf{L}_{N}(S) & \le\mathcal{F}^{\mathsf{Gen}}_{1}.
\end{align}
Substituting these into Eq.~\eqref{eq:apd_max_ineq} yields the fundamental
single-step iterative inequality for the maximal QFI: 
\begin{equation}
\mathcal{F}^{\mathsf{Gen}}_{N}\le\left(\sqrt{\mathcal{F}^{\mathsf{Gen}}_{N-1}}+\sqrt{\mathcal{F}^{\mathsf{Gen}}_{1}}\right)^{2}.\label{eq:apd_iter_ineq}
\end{equation}
To find a closed-form bound, we solve this recurrence inequality.
Let $a_{n}=\mathcal{F}^{\mathsf{Gen}}_{n}$ and $c=\sqrt{\mathcal{F}^{\mathsf{Gen}}_{1}}$.
The inequality is $a_{n}\le(\sqrt{a_{n-1}}+c)^{2}$. By defining a
new sequence $x_{n}=\sqrt{a_{n}}$, we have $x_{n}\le x_{n-1}+c$.
Unfolding this recurrence gives $x_{n}\le x_{1}+(n-1)c=nc$. Squaring
both sides yields the closed-form upper bound $a_{n}\le(nc)^{2}=n^{2}c^{2}$.
We thus obtain: 
\begin{equation}
\mathcal{F}^{\mathsf{Gen}}_{N}\le N^{2}\mathcal{F}^{\mathsf{Gen}}_{1}.
\end{equation}
The maximal single-channel QFI, $\mathcal{F}^{\mathsf{Gen}}_{1}$,
is equivalent to optimizing over all possible input states for a single
channel use, which is known to be $\mathcal{F}^{\mathsf{Gen}}_{1}=\mathcal{F}^{\mathsf{Para}}_{1}=4\min_{\{K_{i}\}}\|\alpha\|$.
Substituting this final piece, we arrive at precisely the HL bound:
\begin{equation}
\mathcal{F}^{\mathsf{Gen}}_{N}(\mathcal{E}_{g})\le4N^{2}\min_{\{K_{i}\}}\|\alpha\|.
\end{equation}
This completes the proof of the universal HL bound.

\section{Global Optimality via KKT Conditions and Alternative Proof of Theorem
1}

\label{sec:kkt}

In this section, we provide a constructive proof of Corollary 1 (Unitary
Equivalence) by solving the convex optimization problem subject to
ICO constraints using the Karush-Kuhn-Tucker (KKT) conditions. By
explicitly identifying the dual variables and verifying global optimality,
we rigorously confirm that for unitary channels, the optimal parallel
strategy is indeed the global optimum across all strategies in $\mathsf{Gen}$.
Furthermore, we demonstrate how this result, combined with the Stinespring
dilation theorem, provides an alternative derivation of the universal
bound stated in Theorem 1. We begin by restating the result to be
proven and giving a sketch of the proof:

\textbf{Corollary 1 (Unitary Equivalence)} For the estimation of a
parameter $g$ encoded in a unitary channel $U_{g}$, the maximal
quantum Fisher information (QFI) achievable with any indefinite causal
order (ICO) strategy is exactly equal to that of the optimal parallel
strategy: 
\begin{equation}
\mathcal{F}^{\mathsf{Gen}}_{N}(U_{g})=\mathcal{F}^{\mathsf{Para}}_{N}(U_{g})=N^{2}(\lambda_{\max}-\lambda_{\min})^{2},
\end{equation}
where $\lambda_{\max}$ and $\lambda_{\min}$ are the maximum and
minimum eigenvalues of the generator $H_{g}=i\dot{U}^{\dagger}_{g}U_{g}$.

\textit{Proof Sketch---}For a unitary channel, the inner minimization
in the QFI expression (Eq.~\eqref{eq:straqfi} of the main text)
can be solved analytically, reducing the problem to maximizing a concave
function over the convex set of ICO strategies $\mathsf{Gen}$. We
propose the known optimal parallel strategy $S^{\star}=\sigma^{\star}\otimes\mathbb{I}_{T^{\mathrm{o}}_{\text{tot}}}$
as a candidate solution~\citep{pang2014QuantumMetrologyGeneral}.
The core of our proof is to demonstrate that this simple strategy
satisfies the Karush-Kuhn-Tucker (KKT) conditions for the much larger
and more complex ICO optimization problem. The KKT conditions require
the existence of dual variables $(E,Y,\mu)$ such that primal and
dual feasibility, complementary slackness, and stationarity are all
satisfied. The stationarity condition links the dual variables via
the relation $Y=\mu\mathbb{I}+\mathcal{Q}^{\dagger}_{\mathsf{Gen}}(E)-\psi$,
where $\psi=A-2\gamma B+\gamma^{2}F$ with $\gamma=N(\lambda_{\mathrm{max}}+\lambda_{\mathrm{min}})/2$.
The operators are defined as $A=|\dot{V}^{\dagger}\rangle\langle\dot{V}^{\dagger}|$,
$B=\frac{1}{2i}(|V^{\dagger}\rangle\langle\dot{V}^{\dagger}|-\mathrm{h.c.})$,
and $F=|V^{\dagger}\rangle\langle V^{\dagger}|$, utilizing the notation
$V=U^{\otimes N}_{g}$. The main challenge lies in constructing a
dual operator $E$ such that the resulting $Y$ is positive semidefinite
($Y\succeq0$) and satisfies complementary slackness ($\mathrm{tr}(YS^{\star})=0$).
We achieve this by introducing a novel recursive construction for
$E$. Assuming a valid set of dual variables $(E^{(n)},\mu^{(n)})$
is known for an $n$-channel problem, we construct the solution for
a $kn$-channel problem. Our ansatz for $E^{(kn)}$ is a symmetrized
operator built from a ``seed'' operator $G^{(kn)}$ that combines
the $n$-channel solution. The seed operator is defined as 
\begin{equation}
G^{(kn)}=(\mathcal{I}-\mathcal{D})\left[\left(\mu^{(n)}\mathbb{I}^{(n)}+E^{(n)}\right)\otimes F^{((k-1)n)}\right],
\end{equation}
where $F$ is the rank-one operator from the QFI expression and $\mathcal{D}$
is the depolarizing superoperator. The full dual variable $E^{(kn)}$
is then constructed by symmetrizing this seed across all $k$ blocks
of channels: 
\begin{equation}
E^{(kn)}=k\sum^{k}_{i=1}\mathrm{S}^{a_{1}a_{i}}G^{(kn)}(\mathrm{S}^{a_{1}a_{i}})^{\dagger},\label{eq:Ekn_maintext}
\end{equation}
where $\mathrm{S}^{a_{1}a_{i}}$ is the swap operator exchanging the
first block of $n$ channels with the $i$-th block. This specific
dual construction for $E$ is designed to ensure that the resulting
operator $Y^{(kn)}$ meets the positivity and complementary slackness
conditions. Since the base case $N=1$ is trivial (as $\mathsf{Gen}$
and $\mathsf{Para}$ sets are identical), this recursive construction
allows us to generate valid dual variables for arbitrary $N$. As
the KKT conditions are sufficient for optimality in this convex problem,
this confirms that the optimal parallel strategy is also the global
optimum across all ICO strategies. The detailed verification is provided
in the following subsections. \hfill{}$\blacksquare$

Having established Corollary 1 via KKT conditions, we can derive Theorem
1 using Stinespring dilation. For any Kraus representation,
let $W_{g}|\psi\rangle=\sum_{i}K_{i}|\psi\rangle|i\rangle$. Then
$W^{\dagger}_{g}W_{g}=\mathbb{I},\;\dot{W}^{\dagger}_{g}\dot{W}_{g}=\alpha,\;W^{\dagger}_{g}\dot{W}_{g}+\dot{W}^{\dagger}_{g}W_{g}=0.$
We first show that a unitary dilation $U_{g}$ can be chosen such
that 
\begin{equation}
H_{g}|\psi,0\rangle=-iU^{\dagger}_{g}\dot{W}_{g}|\psi\rangle,\;\|H_{g}\|^{2}=\|\alpha\|,\label{eq:controlled_unitary_completion}
\end{equation}
where $H_{g}=i\dot{U}^{\dagger}_{g}U_{g}$. Take any unitary dilation
$\bar{U}_{g'}$ satisfying $\bar{U}_{g'}|\psi,0\rangle=W_{g'}|\psi\rangle$,
and define the action of $H_{g}$ on the input subspace by $H_{g}|\psi,0\rangle=-i\bar{U}^{\dagger}_{g}\dot{W}_{g}|\psi\rangle.$
Then 
\begin{align}
\langle\phi,0|H_{g}|\psi,0\rangle & =-i\langle\phi|W^{\dagger}_{g}\dot{W}_{g}|\psi\rangle=\langle\psi,0|H_{g}|\phi,0\rangle^{*},\label{eq:dilation_Hermitian}\\
\sup_{\|\psi\|=1}\|H_{g}|\psi,0\rangle\|^{2} & =\|\dot{W}^{\dagger}_{g}\dot{W}_{g}\|=\|\alpha\|.\label{eq:dilation_norm}
\end{align}
The unspecified block on the orthogonal complement can be chosen so
that $H_{g}$ is Hermitian on the full dilation space and $\|H_{g}\|^{2}=\|\alpha\|$.
Finally, let $\bar{H}_{g}=i\dot{\bar{U}}^{\dagger}_{g}\bar{U}_{g}$.
Since $(H_{g}-\bar{H}_{g})|\psi,0\rangle=0$, define $U_{g'}=\bar{U}_{g'}\exp[i(g'-g)(H_{g}-\bar{H}_{g})].$
It follows directly that 
\begin{equation}
U_{g'}|\psi,0\rangle=W_{g'}|\psi\rangle,\;i\dot{U}^{\dagger}_{g}U_{g}=H_{g},
\end{equation}
which proves Eq.~ \eqref{eq:controlled_unitary_completion}.

Any strategy using $\mathcal{E}_{g}$ is obtained
from a strategy using $U_{g}$ by fixing the environment input and
discarding its output. Corollary 1 therefore gives 
\begin{align}
\mathcal{F}^{\mathsf{Gen}}_{N}(\mathcal{E}_{g}) & \le\mathcal{F}^{\mathsf{Gen}}_{N}(U_{g})\nonumber \\
 & =N^{2}[\lambda_{\max}(H_{g})-\lambda_{\min}(H_{g})]^{2}\nonumber \\
 & \le4N^{2}\|H_{g}\|^{2}=4N^{2}\|\alpha\|.
\end{align}
Minimizing over the Kraus representations proves Eq.~\eqref{eq:bound1}
of the main text and hence Theorem 1.

A detailed proof of Corollary 1 is provided in the following subsections.

\subsection{Derivation of Optimal Parallel Strategies and QFI}

\label{sec:apdparaqfi} In this subsection, we provide a derivation
of the QFI formula for optimal parallel strategies. Let $S\in\mathsf{Para}$
be a parallel strategy and $\sigma:=\mathrm{tr}_{T^{\mathrm{o}}_{\text{tot}}}(S)/d_{\mathrm{out}}$,
where $d_{\mathrm{out}}=d^{N}$. A parallel strategy requires that
$S=\sigma\otimes\mathbb{I}$, where $\sigma$ is therefore the initial
state. With this simplification, the optimization of QFI in Eq.~\eqref{eq:straqfi}
of the main text can be carried out directly. 
\begin{equation}
\begin{aligned}\mathcal{F}^{\mathsf{{Para}}}_{N}(\mathcal{E}_{g}) & =\max_{{S}\in\mathsf{Para}}\min_{h\in\mathbb{R}}\mathrm{tr}(\Omega(h){S})\\
 & =4\max_{\sigma}\left(\operatorname{tr}\left(\sigma G^{2}\right)-\operatorname{tr}\left(\sigma G\right)^{2}\right)\\
 & =(\lambda^{(N)}_{\max}-\lambda^{(N)}_{\min})^{2}=N^{2}(\lambda_{\max}-\lambda_{\min})^{2},
\end{aligned}
\label{eq:parauniqfi}
\end{equation}
where 
\begin{equation}
G:=i\dot{V}^{\dagger}_{g}V_{g}=\sum^{N}_{j=1}H_{j}\label{eq:tothamil}
\end{equation}
is the total generator and $H_{j}=\mathbb{I}^{(j-1)}\otimes H_{g}\otimes\mathbb{I}^{(N-j)}$.
The maximal and minimal eigenvalues of $G$, denoted by $\lambda^{(N)}_{\max}$
and $\lambda^{(N)}_{\min}$, are related to the single-channel eigenvalues
by $\lambda^{(N)}_{\max}=N\lambda_{\max}$ and $\lambda^{(N)}_{\min}=N\lambda_{\min}$.
The extremum over $h$ occurs at $h^{\star}=\mathrm{tr}(SG)/d_{\mathrm{out}}$.
The maximum over $\sigma$ is achieved when its support is restricted
to the subspace spanned by the eigenstates corresponding to the maximal
and minimal eigenvalues of $G$, $\{|e_{M}\rangle,\,|e_{m}\rangle\}$,
and the optimal state $\sigma^{\star}$ takes the form 
\begin{align}
\sigma^{\star}=\frac{1}{2}\begin{pmatrix}1 & x-iy\\
x+iy & 1
\end{pmatrix}\label{eq:paraopti}
\end{align}
in this basis, where $x,y$ are real numbers. Positive-definiteness
requires that $x^{2}+y^{2}\leqslant1$. The optimal strategy is 
\begin{equation}
S^{\star}=\sigma^{\star}\otimes\mathbb{I}.\label{eq:optiS}
\end{equation}
If $x^{2}+y^{2}<1$, the optimal initial state $\sigma^{\star}$ is
mixed and can be purified on a larger system. However, the maximum
QFI can be achieved without ancillary systems by simply choosing $x^{2}+y^{2}=1$,
which makes $\sigma^{\star}$ a pure state. This recovers the well-known
optimal probe state for parallel strategies, an equal superposition
of the eigenstates associated with extremal eigenvalues: $\frac{1}{\sqrt{2}}(|e_{M}\rangle+e^{i\phi}|e_{m}\rangle)$.
Next, we prove the optimality of Eq.~\eqref{eq:optiS}. The analysis
is simplified under the assumption that $G$ is non-degenerate. We
will proceed with this assumption throughout our derivation. The Lagrangian
function is 
\begin{equation}
\mathcal{L}(\sigma,\mu,Y)=\mathrm{tr}(\sigma G^{2})-\mathrm{tr}(\sigma G)^{2}+\mathrm{tr}(Y\sigma)-\mu(\mathrm{tr}(\sigma)-1).
\end{equation}
It satisfies 
\begin{equation}
\min_{Y\succeq0,\mu}\mathcal{L}(\sigma,Y,\mu)=\begin{cases}
\operatorname{var}(\sigma), & \sigma\succeq0,\operatorname{tr}(\sigma)=1.\\
-\infty, & \text{ otherwise }.
\end{cases}
\end{equation}
Therefore, the original problem is rewritten as 
\begin{equation}
\max_{\sigma}\min_{\mu,Y}\mathcal{L}(\sigma,\mu,Y).
\end{equation}
By interchanging the maximization and minimization, we obtain the
dual problem.
\begin{equation}
\min_{\mu,Y}g(\mu,Y)=\min_{\mu,Y}\sup_{\sigma}\mathcal{L}(\sigma,\mu,Y).
\end{equation}
The KKT conditions are
\begin{itemize}
\item Primal feasibility: $\sigma\succeq0,\ \mathrm{tr}(\sigma)=1$.
\item Dual feasibility: $Y\succeq0$.
\item Stationarity: $(G^{2}-2\,\mathrm{tr}(G\sigma)G)+Y-\mu\,\mathbb{I}=0$.
\item Complementary slackness: $\sigma Y=0$.
\end{itemize}
Based on the KKT conditions, the following system of equations is
obtained,
\begin{equation}
\begin{cases}
\text{tr}(G^{2}\sigma)-2\text{tr}(G\sigma)^{2}-\mu=0,\\
(G^{2}-2\text{tr}(G\sigma)G-\mu\mathbb{I})\sigma=0.
\end{cases}\label{eq:KKTcond_1}
\end{equation}
Here, $Y=-\bigl(G^{2}-2\,\mathrm{tr}(G\sigma)G-\mu\,\mathbb{I}\bigr)$.
Performing an eigenvalue decomposition of $G$ yields 
\begin{align}
\sum(\lambda^{(N)2}_{i}-2\text{tr}(G\sigma)\lambda^{(N)}_{i}-\mu)|e_{i}\rangle\langle e_{i}|\sigma=0.
\end{align}
Since $Y$ and $\sigma$ are both positive semidefinite matrices and
their product is zero, we can infer that their support sets are mutually
orthogonal, i.e., 
\begin{align}
\operatorname{supp}\sigma\perp\operatorname{supp}Y.
\end{align}
This is equivalent to the support of $\sigma$ being entirely contained
in the kernel of $Y$, that is, 
\begin{align}
\operatorname{supp}\sigma\subseteq\text{ker}Y.
\end{align}
The kernel of $Y$ is spanned by all eigenvectors $|e_{i}\rangle$
satisfying $\lambda^{(N)2}_{i}-2\text{tr}(G\sigma)\lambda^{(N)}_{i}-\mu=0$.
Since a quadratic equation can have at most two distinct roots, we
conclude that the dimension of the kernel of $Y$ is at most two,
i.e., 
\begin{align}
\text{dim ker}Y\leq2.
\end{align}
According to Vieta's theorem, 
\begin{align}
\begin{cases}
\lambda^{(N)}_{1}+\lambda^{(N)}_{2}=2\text{tr}(G\sigma),\\
\lambda^{(N)}_{1}\lambda^{(N)}_{2}=-\mu.
\end{cases}
\end{align}
By factorizing the matrix $Y$, we obtain 
\begin{align}
Y=(G-\lambda^{(N)}_{1}\mathbb{I})(\lambda^{(N)}_{2}\mathbb{I}-G).\label{eq:ypara_0}
\end{align}
Since $Y$ is a positive semidefinite matrix, we deduce that $\lambda^{(N)}_{1}$
and $\lambda^{(N)}_{2}$ must be the minimum and maximum eigenvalues
of the matrix $G$. If $G$ is non-degenerate, this indicates that
the support of $\sigma$ is confined to the subspace spanned by $|e_{M}\rangle$
and $|e_{m}\rangle$, i.e., 
\begin{align}
\operatorname{supp}\sigma\subseteq\text{span}\{|e_{M}\rangle,\,|e_{m}\rangle\}.
\end{align}
To simplify the analysis, we assume that in the basis $\{|e_{M}\rangle,|e_{m}\rangle\}$,
the matrix representation of $\sigma$ is given by 
\begin{align}
\sigma=\frac{1}{2}\begin{pmatrix}1+z & x-iy\\
x+iy & 1-z
\end{pmatrix}.
\end{align}
By solving the system of equations~\eqref{eq:KKTcond_1}, we obtain
$\lambda_{\max}=\lambda_{\min}$ or $z=0$. This implies that the
matrix representation of $\sigma$ can be further simplified to: 
\begin{align}
\sigma=\frac{1}{2}\begin{pmatrix}1 & x-iy\\
x+iy & 1
\end{pmatrix}.
\end{align}
Finally, we verify that this form of $\sigma$ indeed satisfies all
KKT conditions and derive the upper bound of the variance as 
\begin{align}
\text{Var}(\sigma)=\frac{N^{2}(\lambda_{\max}-\lambda_{\min})^{2}}{4}.
\end{align}
In this example, Slater's condition is satisfied, making the KKT conditions
equivalent to the conditions for the extremum point. An extremum is
attained if and only if $\sigma$ assumes the above form. In summary,
the optimal QFI for the parallel strategy is 
\begin{equation}
\mathcal{F}^{\mathsf{{Para}}}_{N}(\sigma)=N^{2}(\lambda_{\max}-\lambda_{\min})^{2}.
\end{equation}

\subsection{Global Optimality via KKT Conditions}

\label{section:kkt}

In this subsection, we provide the detailed proof that the optimal
parallel strategy $S^{\star}$ satisfies the Karush-Kuhn-Tucker (KKT)
conditions for the more general ICO optimization problem. This verification
is the core of our main result. As stated in the main text, the existence
of valid dual variables $(E,Y,\mu)$ for any number of channels $N$
is proven via a recursive construction. We will show that if a valid
solution to the KKT conditions exists for an $n$-channel problem,
we can construct a valid solution for a $kn$-channel problem for
any integer $k$. We first simplify the min-max problem by solving
the inner minimization over the real parameter $h$. Let $V=U^{\otimes N}_{g}$.
We introduce the following operator notations: 
\begin{align}
A & :=|\dot{V}^{\dagger}\rangle\langle\dot{V}^{\dagger}|,\\
B & :=\frac{1}{2i}\left(|V^{\dagger}\rangle\langle\dot{V}^{\dagger}|-|\dot{V}^{\dagger}\rangle\langle{V}^{\dagger}|\right),\\
F & :=|V^{\dagger}\rangle\langle V^{\dagger}|.
\end{align}
The performance operator $\Omega_{g}(h)$ can be written as a quadratic
function of $h$: $\Omega_{g}(h)=4(A-2hB+h^{2}F)$. The QFI expression
from Eq.~\eqref{eq:straqfi} of the main text is therefore $\mathcal{F}=\max_{S}\min_{h}4\operatorname{tr}((A-2hB+h^{2}F)S)$.
The minimization over $h$ is straightforward, yielding the optimal
value 
\begin{equation}
h^{\star}=\mathrm{tr}(SB)/\mathrm{tr}(SF).
\end{equation}
Substituting this back transforms the QFI optimization into the following
maximization problem: 
\begin{equation}
\mathcal{F}^{\mathsf{Gen}}_{N}(\mathcal{E}_{g})=\max_{S\in\mathsf{Gen}}4\left(\operatorname{tr}(AS)-\frac{[\operatorname{tr}(BS)]^{2}}{\operatorname{tr}(FS)}\right).\label{eq:ICOQFIunirep}
\end{equation}
The objective function in the above optimization can be shown to be
concave with respect to $S$. Furthermore, the feasible region $\mathsf{Gen}$
is a convex set, since it is defined by the intersection of the positive-semidefinite
cone and a set of affine constraints. Therefore, Eq.~\eqref{eq:ICOQFIunirep}
is a convex optimization problem. It also satisfies Slater's condition,
as the normalized identity matrix serves as a strictly feasible point.
To solve the optimization problem, we follow the standard procedure
for constrained optimization. We introduce the dual variables: a scalar
$\mu$, a Hermitian operator $E$, and a positive operator $Y$. The
Lagrangian for the problem is 
\begin{equation}
\begin{aligned} & \mathcal{L}(S,E,Y,\mu)=\operatorname{tr}(AS)-\frac{[\operatorname{tr}(BS)]^{2}}{\operatorname{tr}(FS)}\\
\quad & -\mu\left(\operatorname{tr}(S)-d^{N}\right)-\langle E,{\mathcal{Q}_{\mathsf{Gen}}}(S)\rangle+\langle Y,S\rangle.
\end{aligned}
\end{equation}
The terms with multipliers $\mu,E,Y$ correspond to the trace constraint
$\mathrm{tr}(S)=d^{N}$, the projective constraints $\mathcal{Q}_{\mathsf{Gen}}(S)=0$,
and the positivity constraint $S\succeq0$, respectively. The stationary
condition, obtained by setting the gradient of $\mathcal{L}$ with
respect to $S$ to zero, is 
\begin{equation}
\left(A-\frac{2b}{f}B+\frac{b^{2}}{f^{2}}F\right)-\mu\mathbb{I}-{\mathcal{Q}^{\dagger}_{\mathsf{Gen}}}(E)+Y=0,\label{eq:stationary}
\end{equation}
where $b=\mathrm{tr}(SB)$ and $f=\mathrm{tr}(SF)$. Since this is
a convex optimization problem satisfying Slater's condition, the KKT
conditions are both necessary and sufficient for optimality~\citep{boyd2004ConvexOptimization}.
Therefore, our goal is to find dual variables corresponding to the
solution $S^{\star}=\sigma^{\star}\otimes\mathbb{I}$ (Eq.~\eqref{eq:optiS})
such that the KKT conditions are satisfied. The KKT conditions are
as follows:
\begin{description}
\item [{$\bullet$ Primal feasibility}] The solution must satisfy all original
constraints. 
\begin{equation}
{\mathcal{Q}_{\mathsf{Gen}}}(S)=0,\quad\operatorname{tr}(S)=d^{N},\quad S\succeq0.\label{eq:primalfeasibility}
\end{equation}
\item [{$\bullet$ Dual feasibility}] The dual variable $Y$ must be positive.
\begin{equation}
Y\succeq0.\label{eq:Ygeq0}
\end{equation}
\item [{$\bullet$ Stationarity}] The gradient of the Lagrangian must be
zero, as given in Eq.~\eqref{eq:stationary}.
\item [{$\bullet$ Complementary slackness}] Either the primal constraint
is tight (active) and the dual variable is nonzero, or the primal
constraint has slack (is not tight) and the dual variable is zero.
\begin{equation}
\operatorname{tr}\bigl(YS\bigr)=0.\label{eq:complementarityslack}
\end{equation}
\end{description}
Now, we check whether these four conditions are satisfied by the optimal
parallel strategy $S=S^{\star}$ from Eq.~\eqref{eq:optiS}. First,
the primal feasibility of $S^{\star}$ is immediate, as any valid
parallel strategy is, by definition, a valid ICO strategy ($\mathsf{Para}\subset\mathsf{Gen}$).
Next, we examine the constraints that the KKT conditions impose on
the dual variables $(\mu,Y,E)$. Substitution of $S=S^{\star}$ into
the stationarity condition~\eqref{eq:stationary} yields: 
\begin{equation}
Y=\mu\mathbb{I}+{\mathcal{Q}_{\mathsf{Gen}}}(E)-\psi,\label{eq:stationary_1}
\end{equation}
where 
\begin{equation}
\psi:=A-2\gamma B+\gamma^{2}F.\label{eq:psi}
\end{equation}
Here, we have used the hermiticity of the projector, $\mathcal{Q}^{\dagger}_{\mathsf{Gen}}=\mathcal{Q}_{\mathsf{Gen}}$.
The parameter $\gamma:=\mathrm{tr}(\sigma^{\star}G)=(\lambda^{(N)}_{\mathrm{max}}+\lambda^{(N)}_{\mathrm{min}})/2=N(\lambda_{\mathrm{max}}+\lambda_{\mathrm{min}})/2$
represents the average of the maximal and minimal eigenvalues of the
total generator $G$ (Eq.~\eqref{eq:tothamil}). Taking the trace
of the product of $Y$ (Eq.~\eqref{eq:stationary_1}) and $S^{\star}$,
and using the property of $\mathrm{tr}_{T^{\mathrm{o}}_{\mathrm{tot}}}\mathcal{Q}_{\mathsf{Gen}}(E)=0$,
we obtain 
\begin{equation}
\mu=\frac{N^{2}\left(\lambda_{\max}-\lambda_{\min}\right)^{2}}{4d^{N}}.\label{eq:mu}
\end{equation}
With the solution for $\mu$, we proceed to construct the other dual
variables. We show that for any positive integers $k$ and $n$, a
solution to the KKT conditions $(Y^{(kn)},E^{(kn)})$ for $kn$ channels
can be constructed from a known solution $(Y^{(n)},E^{(n)})$ for
$n$ channels. This recursive step demonstrates that the validity
of the KKT conditions at $N=n$ implies their validity at $N=kn$.
By the equivalence between KKT conditions and optimality, this establishes
that the parallel strategy is optimal within the $\mathsf{Gen}$ set
for all $N$. The base case $N=1$ is trivial, as $\mathsf{Gen}=\mathsf{Para}$
for a single channel. For this case, a dual solution satisfying the
KKT conditions is known to exist, with an explicit form provided in
the Supplemental Material~\ref{sec:apdparaqfi}. The recursive construction
is first shown for the doubling case ($k=2$) and then generalized.
Without loss of generality, the construction assumes that $E^{(n)}$
lies in the image space of $\mathcal{Q}^{(n)}$, i.e., it satisfies
$\mathcal{Q}^{(n)}(E^{(n)})=E^{(n)}$, and then requires the resulting
solution to have the same property: 
\begin{equation}
\mathcal{Q}^{(2n)}(E^{(2n)})=E^{(2n)}.\label{eq:EinimageQ}
\end{equation}
The subscript $\mathsf{Gen}$ of $\mathcal{Q}_{\mathsf{Gen}}$ is
omitted for brevity. Suppose the operator $E^{(n)}$ acts on the space
of the first $n$ channels (denoted by subsystem $a$), ${T^{\mathrm{i}}_{a}T^{\mathrm{o}}_{a}}$,
where $T^{\mathrm{i/o}}_{a}=T^{\mathrm{i/o}}_{1}\cdots T^{\mathrm{i/o}}_{n}$.
To describe the $2n$-channel solution, we introduce a second subsystem
$b$ with the same dimensions, ${T^{\mathrm{i}}_{b}T^{\mathrm{o}}_{b}}$,
where $T^{\mathrm{i/o}}_{b}=T^{\mathrm{i/o}}_{n+1}\cdots T^{\mathrm{i/o}}_{2n}$.
The operator $E^{(2n)}$ is thus an operator on the composite system
$T^{\mathrm{i}}_{a}T^{\mathrm{o}}_{a}T^{\mathrm{i}}_{b}T^{\mathrm{o}}_{b}$.
Using Eq.~\eqref{eq:stationary_1} to eliminate $Y$, the main challenge
is to construct an $E^{(2n)}$ that satisfies both the dual feasibility
and complementary slackness. We propose an ansatz for $E^{(2n)}$
in Eq.~\eqref{eq:E2n}: 
\begin{equation}
E^{(2n)}=2\left(G^{(2n)}+\mathrm{S}^{ab}G^{(2n)}\mathrm{S}^{ab}\right).\label{eq:E2n}
\end{equation}
The seed operator $G^{(2n)}$ is built from the $n$-probe solution
as follows: 
\begin{equation}
G^{(2n)}:=(\mathcal{I}-\mathcal{D})\left[\left(\mu^{(n)}\mathbb{I}^{(n)}+E^{(n)}\right)\otimes F^{(n)}\right],\label{eq:G2n}
\end{equation}
where the superoperator $(\mathcal{I}-\mathcal{D})$ extracts the
traceless component of an operator, and $\mathrm{S}^{ab}$ is the
swap operator for the $a$ and $b$ subsystems. This construction
constitutes the core of our proof.

We begin by examining positive definiteness~\eqref{eq:Ygeq0}. From
Eqs.~\eqref{eq:stationary_1} and~\eqref{eq:E2n}, we obtain 
\begin{equation}
\begin{aligned} & Y^{(2n)}=\mu^{(2n)}\mathbb{I}^{(2n)}+E^{(2n)}-\psi^{(2n)}\\
={} & \mu^{(2n)}\mathbb{I}^{(2n)}+E^{(2n)}-\left(\mathbb{I}^{(n)}+\mathrm{S}^{ab}\right)\psi^{(n)}\otimes F^{(n)}\left(\mathbb{I}^{(n)}+\mathrm{S}^{ab}\right)^{\dagger}\\
={} & \left(\mathbb{I}^{(2n)}-\mathrm{S}^{ab}\right)\left(\psi^{(n)}+Y^{(n)}\right)\otimes F^{(n)}\left(\mathbb{I}^{(2n)}-\mathrm{S}^{ab}\right)^{\dagger}\\
 & +\left(\mathbb{I}^{(2n)}+\mathrm{S}^{ab}\right)Y^{(n)}\otimes F^{(n)}\left(\mathbb{I}^{(2n)}+\mathrm{S}^{ab}\right)^{\dagger}.
\end{aligned}
\label{eq:Y2n_1}
\end{equation}
Using $F^{(n)},\psi^{(n)}\succeq0$ (Eq.~\eqref{eq:psi}), the dual
feasibility $Y^{(n)}\succeq0$ from the inductive hypothesis, and
the fact that the map with Kraus operators $\bigl(\mathbb{I}^{(2n)}\pm\mathrm{S}^{ab}\bigr)$
preserves positive definiteness, we obtain 
\begin{equation}
Y^{(2n)}\succeq0.
\end{equation}
Complementary slackness: We strengthen the assumption by supposing
that for $N=n$, $\mathrm{supp}\,Y^{(n)}\subset(\mathrm{span}\{|e_{m}\rangle,|e_{M}\rangle\}\otimes\mathcal{H}^{T^{\mathrm{o}}_{\text{tot}}})^{\perp}$.
That is, for all $|\theta\rangle\in\mathcal{H}^{T^{\mathrm{o}}_{\text{tot}}}$,
$Y^{(n)}|e_{m/M}\rangle\otimes|\theta\rangle=0$. This constitutes
a sufficient condition for the complementary slackness property, since
the $T^{\mathrm{i}}_{\text{tot}}$ factor of $S$ in Eq.~\eqref{eq:optiS}
is supported on $\mathrm{span}\{|e_{m}\rangle,\,|e_{M}\rangle\}$.
It is therefore sufficient to prove the following: 
\begin{equation}
\mathrm{supp}~Y^{(2n)}\subset\left(\mathrm{span}\{|e_{m}\rangle^{(2n)}_{T^{\mathrm{i}}_{\text{tot}}},|e_{M}\rangle^{(2n)}_{T^{\mathrm{i}}_{\text{tot}}}\}\otimes\mathcal{H}^{T^{\mathrm{o}}_{\text{tot}}}\right)^{\perp}.\label{eq:slackness_2n}
\end{equation}
To prove the above relation, it suffices to consider the terms in
Eq.~\eqref{eq:Y2n_1} that do not involve $Y^{(n)}$, namely, 
\begin{equation}
\begin{aligned} & \left(\mathbb{I}^{(2n)}-\mathrm{S}^{ab}\right)\psi^{(n)}\otimes F^{(n)}\left(\mathbb{I}^{(2n)}-\mathrm{S}^{ab}\right)^{\dagger}\\
={} & \left(\mathbb{I}-\mathrm{S}^{ab}\right)(\psi\otimes F+F\otimes\psi).
\end{aligned}
\end{equation}
It suffices to show that, for all $|y\rangle,|z\rangle$, we have
\begin{equation}
\left(\mathbb{I}-\mathrm{S}^{ab}\right)(\psi\otimes F+F\otimes\psi)|e_{m}\rangle\otimes|y\rangle\otimes|e_{m}\rangle\otimes|z\rangle=0,\label{eq:slackness_1}
\end{equation}
and 
\begin{equation}
\left(\mathbb{I}-\mathrm{S}^{ab}\right)(\psi\otimes F+F\otimes\psi)|e_{M}\rangle\,\otimes|y\rangle\otimes|e_{M}\rangle\otimes|z\rangle=0.\label{eq:slackness_2}
\end{equation}

From Eq.~\eqref{eq:psi}, we have 
\begin{equation}
\psi=\left[(G-\gamma\mathbb{I})\otimes\mathbb{I}\right]|V^{\dagger}\rangle\langle V^{\dagger}|\left[(G-\gamma\mathbb{I})\otimes\mathbb{I}\right],
\end{equation}
where $V=U^{\otimes N}$. We now proceed to prove Eq.~\eqref{eq:slackness_1}:

\begin{equation}
\begin{aligned} & (\mathbb{I}-\mathrm{S}^{ab})(\psi\otimes F+F\otimes\psi)|e_{m}\rangle\otimes|y\rangle\otimes|e_{m}\rangle\otimes|z\rangle\\
={} & (\mathbb{I}-\mathrm{S}^{ab})\Bigg\{\Bigl[((G-\gamma\mathbb{I})\otimes\mathbb{I})|V^{\dagger}\rangle\langle V^{\dagger}|((G-\gamma\mathbb{I})\otimes\mathbb{I})\Bigr]\otimes|V^{\dagger}\rangle\langle V^{\dagger}|\\
 & \qquad+|V^{\dagger}\rangle\langle V^{\dagger}|\otimes\Bigl[((G-\gamma\mathbb{I})\otimes\mathbb{I})|V^{\dagger}\rangle\langle V^{\dagger}|((G-\gamma\mathbb{I})\otimes\mathbb{I})\Bigr]\Bigg\}\\
 & \qquad\qquad\times|e_{m}\rangle\otimes|y\rangle\otimes|e_{m}\rangle\otimes|z\rangle\\
={} & (\lambda_{\min}-\gamma)(\mathbb{I}-\mathrm{S}^{ab})\Bigg\{\Bigl[((G-\gamma\mathbb{I})\otimes\mathbb{I})|V^{\dagger}\rangle\langle V^{\dagger}|(|e_{m}\rangle\otimes|y\rangle)\Bigr]\otimes\Bigl[|V^{\dagger}\rangle\langle V^{\dagger}|(|e_{m}\rangle\otimes|z\rangle)\Bigr]\\
 & \qquad+\Bigl[|V^{\dagger}\rangle\langle V^{\dagger}|(|e_{m}\rangle\otimes|y\rangle)\Bigr]\otimes\Bigl[((G-\gamma\mathbb{I})\otimes\mathbb{I})|V^{\dagger}\rangle\langle V^{\dagger}|(|e_{m}\rangle\otimes|z\rangle)\Bigr]\Bigg\}\\
={} & (\lambda_{\min}-\gamma)\langle V^{\dagger}|(|e_{m}\rangle\otimes|y\rangle)\langle V^{\dagger}|(|e_{m}\rangle\otimes|z\rangle)\Bigg\{((G-\gamma\mathbb{I})\otimes\mathbb{I})|V^{\dagger}\rangle\otimes|V^{\dagger}\rangle\\
 & \qquad+|V^{\dagger}\rangle\otimes((G-\gamma\mathbb{I})\otimes\mathbb{I})|V^{\dagger}\rangle-|V^{\dagger}\rangle\otimes((G-\gamma\mathbb{I})\otimes\mathbb{I})|V^{\dagger}\rangle\\
 & \qquad-((G-\gamma\mathbb{I})\otimes\mathbb{I})|V^{\dagger}\rangle\otimes|V^{\dagger}\rangle\Bigg\}\\
={} & 0.
\end{aligned}
\end{equation}

The first equality follows from the definitions of $\psi$ and $F$.
The second equality uses $G|e_{m}\rangle=\lambda_{\min}|e_{m}\rangle$
to evaluate the rightmost factor $(G-\gamma\mathbb{I})\otimes\mathbb{I}$
on the input state in the corresponding block. The third equality
uses the action of $(\mathbb{I}-\mathrm{S}^{ab})$. The fourth equality
uses the rank-one form of $F$ and expands the action of the swap.
Similarly, Eq.~\eqref{eq:slackness_2} holds. Hence, Eq.~\eqref{eq:slackness_2n}
holds. Finally, it must be verified that the constructed operator
$E^{(2n)}$ lies in the image of the projector $\mathcal{Q}^{(2n)}$:
\begin{equation}
E^{(2n)}=2\left(G^{(2n)}+\mathrm{S}^{ab}G^{(2n)}\mathrm{S}^{ab}\right)\in\text{im}\mathcal{Q}^{(2n)}.
\end{equation}
Since the swap superoperator $\mathcal{S}^{ab}(\cdot):=\mathrm{S}^{ab}(\cdot)\mathrm{S}^{ab}$
commutes with $\mathcal{Q}^{(2n)}$, it is sufficient to show that
$G^{(2n)}\in\mathrm{im}\,\mathcal{Q}^{(2n)}$. The operator $G^{(2n)}$
can be decomposed as 
\begin{equation}
G^{(2n)}=E^{(n)}\otimes\frac{\mathbb{I}^{(n)}}{d^{n}}+E^{(n)}\otimes\widetilde{F}^{(n)}+\mu^{(n)}\mathbb{I}^{(n)}\otimes\widetilde{F}^{(n)},\label{eq:G2na}
\end{equation}
where $\tilde{F}=(\mathcal{I}-\mathcal{D})F$ denotes the traceless
component of $F$. Each term in this decomposition can be shown to
lie in $\mathrm{im}\,\mathcal{Q}^{(2n)}$. Let $\mathcal{Q}^{(2n)}=\bigotimes^{2n}_{i=1}(\mathcal{I}-\mathcal{D}_{T^{\mathrm{o}}_{i}}+\mathcal{D}_{T_{i}})-\mathcal{D}$,
$\mathcal{Q}^{(n)}\equiv\mathcal{Q}^{a}=\bigotimes^{n}_{i=1}(\mathcal{I}-\mathcal{D}_{T^{\mathrm{o}}_{i}}+\mathcal{D}_{T_{i}})-\mathcal{D}$,
and $\mathcal{Q}^{b}=\bigotimes^{2n}_{i=n+1}(\mathcal{I}-\mathcal{D}_{T^{\mathrm{o}}_{i}}+\mathcal{D}_{T_{i}})-\mathcal{D}$.
For the first term, using the induction hypothesis $\mathcal{Q}^{(n)}(E^{(n)})=E^{(n)}$
and the fact that $E^{(n)}$ is traceless, we obtain 
\begin{equation}
\begin{aligned} & \mathcal{Q}^{(2n)}\left(E^{(n)}\otimes\mathbb{I}^{(n)}\right)\\
={} & \bigotimes^{2n}_{i=1}\left(\mathcal{I}-\mathcal{D}_{T^{\mathrm{o}}_{i}}+\mathcal{D}_{T_{i}}\right)\left(E^{(n)}\otimes\mathbb{I}^{(n)}\right)\\
={} & \bigotimes^{n}_{i=1}\left(\mathcal{I}-\mathcal{D}_{T^{\mathrm{o}}_{i}}+\mathcal{D}_{T_{i}}\right)\left(E^{(n)}\otimes\mathbb{I}^{(n)}\right)=E^{(n)}\otimes\mathbb{I}^{(n)}.
\end{aligned}
\end{equation}
Similar calculations, omitted here, confirm that the other terms in
Eq.~\eqref{eq:G2na} also lie in the image. Notably, this proof relies
only on general properties of the projector $\mathcal{Q}_{\mathsf{Gen}}$
(its behavior under tensor products and its permutation symmetry).
Specifically, we rely on the fact that if two subsystem operators,
$E^{A}\in\mathrm{im}\mathcal{Q}^{A}$ and $E^{B}\in\mathrm{im}\mathcal{Q}^{B}$,
then their tensor products with the identity operator lie in $\mathrm{im}\mathcal{Q}^{AB}$,
i.e., $E^{A}\otimes E^{B}\in\mathrm{im}\mathcal{Q}^{AB}$, $E^{A}\otimes\mathbb{I}\in\mathrm{im}\mathcal{Q}^{AB}$,
and $\mathbb{I}\otimes E^{B}\in\mathrm{im}\mathcal{Q}^{AB}$. Another
property is that $\mathcal{Q}$ is symmetric with respect to the exchange
of any two channels. This implies our result holds for any strategy
set whose projector shares these features. Thus, we have presented
the expressions for $(S^{(2n)},E^{(2n)},Y^{(2n)},\mu^{(2n)})$, which
satisfy the KKT conditions. The above procedure can be iterated to
generate solutions for $E^{(2)},E^{(4)},E^{(8)},\dots$, thereby proving
the equivalence between parallel and $\mathsf{Gen}$ strategies for
all $N=2^{m}$ with $m\in\mathbb{Z}^{+}$. To complete the proof for
all positive integers $N$, we generalize the construction to an arbitrary
integer factor $k$. In this scenario, $E^{(kn)}$ acts on a system
composed of $k$ subsystems of $n$ channels each, $T^{\mathrm{i}}_{a_{1}}T^{\mathrm{o}}_{a_{1}}\cdots T^{\mathrm{i}}_{a_{k}}T^{\mathrm{o}}_{a_{k}}$.
The generalization of Eq.~\eqref{eq:E2n} is 
\begin{equation}
E^{(kn)}=k\sum^{k}_{i=1}\mathrm{S}^{a_{1}a_{i}}G^{(kn)}\mathrm{S}^{a_{1}a_{i}},\label{eq:Ekn}
\end{equation}
where 
\begin{equation}
G^{(kn)}=(\mathcal{I}-\mathcal{D})\left[\left(\mu^{(n)}\mathbb{I}^{(n)}+E^{(n)}\right)\otimes F^{((k-1)n)}\right].
\end{equation}
Here, $\mathrm{S}^{a_{1}a_{1}}:=\mathbb{I}$, and $\mathrm{S}^{a_{1}a_{i}}$
is the swap operator between the first and $i$-th blocks~($T^{\mathrm{i}}_{a_{1}}T^{\mathrm{o}}_{a_{1}}\leftrightarrow T^{\mathrm{i}}_{a_{i}}T^{\mathrm{o}}_{a_{i}}$).
This expression correctly reduces to Eq.~\eqref{eq:E2n} for $k=2$.

Next, we prove that for an arbitrary $k$, the tuple $(S^{(kn)},E^{(kn)},Y^{(kn)})$
satisfies the KKT conditions. Introducing the notation $\mathrm{S}^{i}:=\mathrm{S}^{a_{1}a_{i}}$
and using the definitions of $Y$ in Eq.~\eqref{eq:stationary_1}
and $E^{(kn)}$ in Eq.~\eqref{eq:Ekn}, we obtain 
\begin{equation}
\begin{aligned}Y^{(kn)}={} & \mu^{(kn)}\mathbb{I}^{(kn)}+E^{(kn)}-\left(\sum^{k}_{i,j=1}\mathrm{S}^{i}\psi^{(n)}\otimes F^{((k-1)n)}\mathrm{S}^{j}\right)\\
={} & k\sum^{k}_{i=1}\mathrm{S}^{i}\left(Y^{(n)}\otimes F^{((k-1)n)}\right)\mathrm{S}^{i}+\mathcal{S}\!\left(\psi^{(n)}\otimes F^{((k-1)n)}\right),
\end{aligned}
\end{equation}
where 
\begin{equation}
\mathcal{S}(R)=\sum^{k}_{i,j=1}(k\delta_{ij}-1)\mathrm{S}^{i}R\mathrm{S}^{j}.\label{eq:SR}
\end{equation}
The coefficient matrix $(k\delta_{ij}-1)^{k}_{i,j=1}$ is positive
semidefinite. Equivalently, the map $\mathcal{S}$ admits the Kraus
representation 
\begin{equation}
\mathcal{S}(R)=\sum_{1\le i<j\le k}(\mathrm{S}^{i}-\mathrm{S}^{j})R(\mathrm{S}^{i}-\mathrm{S}^{j})^{\dagger}.
\end{equation}
Hence $\mathcal{S}$ is completely positive. Consequently, since $Y^{(n)}\succeq0$
and $F^{((k-1)n)}\succeq0$, we have 
\begin{equation}
Y^{(kn)}\succeq0.
\end{equation}

The proofs of the complementary slackness and $E^{(kn)}\in\mathrm{im}\mathcal{Q}_{\mathsf{Gen}}$
are analogous to those for $k=2$ and are therefore omitted.

\section{Derivation of Structurally Refined Bound}

\label{sec:bound2_derivation}

In this section, we present the rigorous derivation of the structurally
refined upper bound on the QFI for general ICO strategies, as stated
in Theorem 2. The central objective is to prove that for any quantum
channel $\mathcal{E}_{g}$ and any strategy $S\in\mathsf{Gen}$, the
maximal QFI is bounded by 
\begin{equation}
\mathcal{F}^{\mathsf{Gen}}_{N}(\mathcal{E}_{g})\le\min_{\{K_{i}\}}\bigg[4N\|\alpha\|+N(N-1)\left(4\|\beta\|^{2}+2d\|\beta\|\|\Lambda^{\perp}\|\right)\bigg],
\end{equation}
where the minimization is performed over all Kraus representations
$\{K_{i}\}$ of the channel. Our proof employs an iterative construction
of the performance operator, decomposing the QFI into local and nonlocal
contributions and exploiting the orthogonality between interference
operators and reduced process matrices to tighten the bound. The optimization
of the QFI is formally given by $\mathcal{F}^{\mathsf{Gen}}_{N}(\mathcal{E}_{g})=\max_{S}\min_{h}\mathrm{tr}(\Omega^{(N)}(h)S)$.
The performance operator $\Omega^{(N)}(h)$ is defined in terms of
an ensemble decomposition of the total channel's Choi matrix, $\mathsf{E}^{\otimes N}_{g}=\mathbf{C}^{0}_{g}(\mathbf{C}^{0}_{g})^{\dagger}$,
as $\Omega^{(N)}(h)=4((\dot{\mathbf{C}}^{0}_{g}-i\mathbf{C}^{0}_{g}h)(\dot{\mathbf{C}}^{0}_{g}-i\mathbf{C}^{0}_{g}h)^{\dagger})^{\top}$.
A particularly important choice for the reference ensemble is one
constructed from the tensor products of single-channel Kraus operators
$\{K_{i}\}$. Specifically, each column of the ensemble matrix $\mathbf{C}^{0}_{g}$
corresponds to a vectorized operator $|\mathbf{K}^{\top}_{\mathbf{i}}\rangle$
where $\mathbf{K}_{\mathbf{i}}:=K_{i_{1}}\otimes\cdots\otimes K_{i_{N}}$
and the index $\mathbf{i}=(i_{1},\dots,i_{N})$ is a multi-index.
The performance operator takes the form $\Omega^{(N)}(0)=4\sum_{\mathbf{i}}|\dot{\mathbf{K}}^{\dagger}_{\mathbf{i}}\rangle\langle\dot{\mathbf{K}}^{\dagger}_{\mathbf{i}}|$.
The minimization over the parameter $h$ is encompassed by the freedom
in choosing the Kraus representation of $\mathcal{E}^{\otimes N}_{g}$.
The recursive structure of the multi-channel performance operator
is governed by the single-channel interference operator $\Lambda$,
defined on the joint input-output space $T^{\mathrm{i}}\otimes T^{\mathrm{o}}$
as $\Lambda:=4\sum_{i}|\dot{K}^{\dagger}_{i}\rangle\langle K^{\dagger}_{i}|$.
The magnitude of the interference effects, and thus the potential
for Heisenberg scaling, depends on how this operator $\Lambda$ correlates
the input and output spaces. To precisely quantify this, we introduce
an orthogonal decomposition of $\Lambda$. We separate it into a ``parallel''
component, which projects onto the identity of the output space, and
an ``orthogonal'' component. The parallel component $\Lambda^{\parallel}$
is defined as the part of $\Lambda$ that contains $\beta$: 
\begin{equation}
\Lambda^{\parallel}:=\mathcal{D}_{T^{\mathrm{o}}}\Lambda=4\beta\otimes\frac{\mathbb{I}_{\mathrm{out}}}{d},
\end{equation}
where $d=\dim(T^{\mathrm{o}})$. The orthogonal component $\Lambda^{\perp}$
is defined as the remainder of the interference operator, $\Lambda^{\perp}:=\Lambda-\Lambda^{\parallel}$.
This decomposition is structurally significant because $\Lambda^{\perp}$
is traceless on the output subsystem. Explicitly, we have 
\begin{equation}
\mathrm{tr}_{\mathrm{out}}(\Lambda^{\perp})=\mathrm{tr}_{\mathrm{out}}(\Lambda)-\mathrm{tr}_{\mathrm{out}}\left(4\beta\otimes\frac{\mathbb{I}_{\mathrm{out}}}{d}\right)=0.\label{eq:traceless_property}
\end{equation}
This property will be pivotal in eliminating specific nonlocal terms
later in the proof. To evaluate the QFI for $N$ channels, we require
an explicit expression for the total performance operator $\Omega^{(N)}$.
Following the recursive construction for quantum channels developed
in Ref.~\citep{fujiwara2008FibreBundleManifolds}, $\Omega^{(N)}$
can be expanded exactly into a sum of single-channel local terms and
pairwise inter-channel interference terms. Let $\Omega^{(1)}=4\sum_{i}|\dot{K}^{\dagger}_{i}\rangle\langle\dot{K}^{\dagger}_{i}|$
denote the single-channel performance operator and $\mathsf{E}$ the
Choi matrix of the channel. The total performance operator acts on
the tensor product space of $N$ inputs and $N$ outputs. Its expansion
is given by $\Omega^{(N)}=\Omega^{(N)}_{\text{local}}+\Omega^{(N)}_{\text{nonlocal}}$.
The local contribution represents the incoherent accumulation of precision:
\begin{equation}
\Omega^{(N)}_{\text{local}}=\sum^{N}_{k=1}(\mathsf{E}^{\top})^{\otimes(k-1)}\otimes\Omega^{(1)}\otimes(\mathsf{E}^{\top})^{\otimes(N-k)},
\end{equation}
where the notation implies that $\Omega^{(1)}$ acts on the $k$-th
channel subspace while the Choi matrices act on the others. The nonlocal
contribution captures the coherent interference between all pairs
of channels.

\begin{equation}
\Omega^{(N)}_{\text{nonlocal}}=\frac{1}{4}\sum_{1\le j<k\le N}\left[(\mathsf{E}^{\top})^{\otimes(j-1)}\otimes\Lambda\otimes(\mathsf{E}^{\top})^{\otimes(k-j-1)}\otimes\Lambda^{\dagger}\otimes(\mathsf{E}^{\top})^{\otimes(N-k)}+\text{h.c.}\right].\label{eq:Omega_expansion_nonlocal}
\end{equation}

Consequently, the total QFI for a strategy $S$ decomposes as $\mathcal{F}_{N}(S)=\mathrm{tr}(\Omega^{(N)}_{\text{local}}S)+\mathrm{tr}(\Omega^{(N)}_{\text{nonlocal}}S)$.
We now analyze the nonlocal term by substituting the decomposition
$\Lambda=\Lambda^{\parallel}+\Lambda^{\perp}$. For any pair of channels
$(j,k)$, the core interference term $\Lambda_{j}\otimes\Lambda^{\dagger}_{k}$
expands into four distinct components: 
\begin{equation}
\Lambda_{j}\otimes\Lambda^{\dagger}_{k}={\Lambda^{\parallel}_{j}\otimes\Lambda^{\parallel\dagger}_{k}}+{\Lambda^{\parallel}_{j}\otimes\Lambda^{\perp\dagger}_{k}}+{\Lambda^{\perp}_{j}\otimes\Lambda^{\parallel\dagger}_{k}}+{\Lambda^{\perp}_{j}\otimes\Lambda^{\perp\dagger}_{k}}.\label{eq:four_terms}
\end{equation}

\textbf{Lemma 1} For any valid $N$-party strategy $S\in\mathsf{Gen}$,
and any pair of channels $(j,k)$, the contribution to the QFI from
the term involving only orthogonal components vanishes: 
\begin{equation}
\mathrm{tr}\left[\left(\Lambda^{\perp}_{j}\otimes(\Lambda^{\perp}_{k})^{\dagger}\right)S^{\mathrm{red}}_{jk}\right]=0,
\end{equation}
where 
\begin{equation}
S^{\mathrm{red}}_{jk}=S\star((\mathsf{E}^{\top})^{\otimes(j-1)}\otimes(\mathsf{E}^{\top})^{\otimes(k-j-1)}\otimes(\mathsf{E}^{\top})^{\otimes(N-k)})\label{eq:reduced_two_use_process}
\end{equation}
is the reduced two-party process matrix. \textit{Proof.} The physical
validity of a strategy $S$ is defined by the constraint $\mathcal{Q}_{\mathsf{Gen}}(S)=0$,
meaning $S$ lies in the kernel of the ICO projector. Consequently,
any reduced process $S^{\mathrm{red}}_{jk}$ must also satisfy the
corresponding two-party constraints. The image space of the projector
$\mathrm{im}(\mathcal{Q}_{\mathsf{Gen}})$ consists of all operators
that violate these causal consistency conditions. Recall from Eq.~\eqref{eq:traceless_property}
that $\Lambda^{\perp}$ is constructed to be traceless on the output
space. Therefore, the tensor product $X_{\perp\perp}:=\Lambda^{\perp}_{j}\otimes(\Lambda^{\perp}_{k})^{\dagger}$
is traceless on both $T^{\mathrm{o}}_{j}$ and $T^{\mathrm{o}}_{k}$.
This structural property guarantees that $X_{\perp\perp}\in\mathrm{im}(\mathcal{Q}_{\mathsf{Gen}})$.
Since the physically valid strategy $S^{\mathrm{red}}_{jk}$ resides
in the kernel of $\mathcal{Q}_{\mathsf{Gen}}$, and the kernel and
image spaces are orthogonal, their inner product must be zero: $\mathrm{tr}(X_{\perp\perp}S^{\mathrm{red}}_{jk})=0$.
\hfill{}$\blacksquare$

With the $(\perp,\perp)$ term eliminated, we proceed to bound the
remaining terms in the QFI expansion. We treat the local and nonlocal
contributions separately using the triangle inequality and the Hölder
inequality $|\mathrm{tr}(AB)|\le\|A\|_{\infty}\|B\|_{1}$. First,
consider the local contribution $\mathrm{tr}(\Omega^{(N)}_{\text{local}}S)$.
The local term scales linearly with $N$ and is bounded by the standard
single-channel limit: 
\begin{equation}
\left|\mathrm{tr}(\Omega^{(N)}_{\text{local}}S)\right|\le4N\|\alpha\|.
\end{equation}

Next, we bound the surviving nonlocal interference terms. We sum over
all $N(N-1)/2$ pairs of channels. For a given pair $(j,k)$, the
reduced process matrix $S^{\mathrm{red}}_{jk}$ satisfies the normalization
condition $\|S^{\mathrm{red}}_{jk}\|_{1}=\mathrm{tr}(S^{\mathrm{red}}_{jk})=d^{2}$,
where $d$ is the dimension of the single-channel output space. The
first surviving nonlocal term is the parallel-parallel interaction,
$(\Lambda^{\parallel}_{j}\otimes(\Lambda^{\parallel}_{k})^{\dagger}+\text{h.c.})$.
The operator norm of the parallel component is $\|\Lambda^{\parallel}\|=\|4\beta\otimes(\mathbb{I}/d)\|=4\|\beta\|/d$.
Applying the Hölder inequality: 
\begin{equation}
\left|\frac{1}{4}\mathrm{tr}\left[\left(\Lambda^{\parallel}_{j}\otimes(\Lambda^{\parallel}_{k})^{\dagger}+\text{h.c.}\right)S^{\mathrm{red}}_{jk}\right]\right|\le\frac{1}{4}\cdot2\cdot\|\Lambda^{\parallel}\|^{2}d^{2}=8\|\beta\|^{2}.
\end{equation}
Summing this over all pairs yields a total contribution of $\frac{N(N-1)}{2}\times8\|\beta\|^{2}=4N(N-1)\|\beta\|^{2}$.
This term recovers the standard Heisenberg scaling bound for parallel
strategies, confirming that our decomposition is consistent with known
limits. The remaining nonlocal terms are the cross-couplings between
the parallel and the orthogonal components: $(\Lambda^{\parallel}_{j}\otimes(\Lambda^{\perp}_{k})^{\dagger}+\Lambda^{\perp}_{j}\otimes(\Lambda^{\parallel}_{k})^{\dagger}+\text{h.c.})$.
There are four such terms in the Hermitian expansion. The norm of
each is bounded by $\|\Lambda^{\parallel}\|\|\Lambda^{\perp}\|d^{2}$.
Summing over all pairs yields $N(N-1)\cdot2d\|\beta\|\|\Lambda^{\perp}\|$.
Finally, by combining the bounds for the local term, the parallel-parallel
interference, and the parallel-orthogonal interference, we obtain
an upper bound for the QFI for a specific Kraus representation. We
minimize over all possible Kraus decompositions $\{K_{i}\}$ to obtain
the tightest limit. This leads to the final derivation of Theorem
2. The entire derivation, from the formal definition of the QFI to
the final bound, can be summarized in the following sequence of relations.

\begin{align}
\mathcal{F}^{\mathsf{Gen}}_{N}(\mathcal{E}_{g}) & =\max_{S\in\mathsf{Gen}}\min_{h\in\mathbb{H}_{r^{N}}}\mathrm{tr}\left(\Omega^{(N)}(h)S\right)=\max_{S\in\mathsf{Gen}}\min_{\{\mathbf{K}_{\mathbf{i}}\}}\mathrm{tr}\left(4\sum_{\mathbf{i}}|\dot{\mathbf{K}}^{\dagger}_{\mathbf{i}}\rangle\langle\dot{\mathbf{K}}^{\dagger}_{\mathbf{i}}|S\right)\label{eq:final_derivation_1}\\
 & \le\min_{\{K_{j}\}}\max_{S\in\mathsf{Gen}}\mathrm{tr}\left(\Omega^{(N)}_{\{K_{j}\}}S\right)\nonumber \\
 & =\min_{\{K_{j}\}}\max_{S\in\mathsf{Gen}}\Bigg\{\mathrm{tr}\left[\left(\sum^{N}_{k=1}(\mathsf{E}^{\top})^{\otimes(k-1)}\otimes\Omega^{(1)}\otimes(\mathsf{E}^{\top})^{\otimes(N-k)}\right)S\right]+\frac{1}{4}\sum_{1\le j<k\le N}\mathrm{tr}\left[\left(\Lambda_{j}\otimes\Lambda^{\dagger}_{k}+\text{h.c.}\right)S^{\mathrm{red}}_{jk}\right]\Bigg\}\label{eq:final_derivation_2}\\
 & =\min_{\{K_{j}\}}\max_{S\in\mathsf{Gen}}\Bigg\{\mathrm{tr}\left(\sum^{N}_{k=1}\Omega^{(1)}S^{\mathrm{red}}_{k}\right)+\frac{1}{4}\sum_{1\le j<k\le N}\mathrm{tr}\left[\left((\Lambda^{\parallel}_{j}+\Lambda^{\perp}_{j})\otimes(\Lambda^{\parallel}_{k}+\Lambda^{\perp}_{k})^{\dagger}+\text{h.c.}\right)S^{\mathrm{red}}_{jk}\right]\Bigg\}\label{eq:final_derivation_3}\\
 & \le\min_{\{K_{j}\}}\left[4N\|\alpha\|+N(N-1)\left(4\|\beta\|^{2}+2d\|\beta\|\|\Lambda^{\perp}\|\right)\right].\label{eq:final_derivation_4}
\end{align}

The derivation begins in Eq.~\eqref{eq:final_derivation_1} by stating
that the minimization over the parameter $h$ is equivalent to minimizing
over all global Kraus representations $\{\mathbf{K}_{\mathbf{i}}\}$
for the $N$-channel process. To obtain a tractable upper bound, we
restrict this minimization to product-form Kraus representations $\{K_{j}\}^{\otimes N}$
derived from a single-channel decomposition. In Eq.~\eqref{eq:final_derivation_2},
the resulting performance operator is expanded into its local and
nonlocal parts, where $S^{\mathrm{red}}_{jk}$ is the reduced two-party
process matrix. In Eq.~\eqref{eq:final_derivation_3}, we substitute
the orthogonal decomposition of $\Lambda$ into its parallel and orthogonal
components. The crucial step to reach Eq.~\eqref{eq:final_derivation_4}
is the application of a structural lemma: the term involving purely
orthogonal correlations, $\mathrm{tr}[(\Lambda^{\perp}_{j}\otimes\Lambda^{\perp\dagger}_{k})S^{\mathrm{red}}_{jk}]$,
vanishes for any valid ICO strategy. This is because the operator
$\Lambda^{\perp}_{j}\otimes\Lambda^{\perp\dagger}_{k}$ lies in the
image space of the ICO projector $\mathcal{Q}_{\mathsf{Gen}}$, while
any valid strategy $S$ resides in its kernel, making their contraction
zero. We then apply the Hölder inequality to the surviving terms---local,
parallel-parallel, and cross-terms---and sum over all $N(N-1)/2$
pairs to obtain the final inequality. This bound explicitly isolates
the scaling behavior. For channels where the Hamiltonian lies in the
span of the Kraus operators, there exists a representation where $\beta=0$.
In such cases, both the parallel-parallel term and the cross-coupling
term vanish, forcing the $O(N^{2})$ coefficient to zero. This rigorously
proves that ICO strategies cannot alter the fundamental SQL scaling
for this class of noisy channels. Finally, we demonstrate that the
orthogonal component $\Lambda^{\perp}$ is non-vanishing for any non-trivial
parameter estimation problem. By definition, the condition $\Lambda^{\perp}=0$
implies that the signal is purely parallel, which translates to the
algebraic constraint on the Kraus operators: 
\begin{equation}
\sum_{i}|\dot{K}^{\dagger}_{i}\rangle\langle K^{\dagger}_{i}|=\left(\sum_{j}\dot{K}^{\dagger}_{j}K_{j}\right)\otimes\frac{\mathbb{I}}{d}.
\end{equation}
To see the physical implication of this, we examine the derivative
of the channel's Choi matrix, $\mathsf{E}^{\top}=\sum_{i}|K^{\dagger}_{i}\rangle\langle K^{\dagger}_{i}|$.
Applying the Leibniz rule and substituting the condition above (along
with its Hermitian conjugate), we find that the derivative takes the
form $\dot{\mathsf{E}}^{\top}=(\beta+\beta^{\dagger})\otimes(\mathbb{I}/d)$,
where $\beta=\sum_{i}\dot{K}^{\dagger}_{i}K_{i}$. However, the trace-preserving
condition of the quantum channel, $\sum_{i}K^{\dagger}_{i}K_{i}=\mathbb{I}$,
necessitates that $\beta$ is anti-Hermitian (i.e., $\beta+\beta^{\dagger}=0$).
Consequently, the assumption $\Lambda^{\perp}=0$ leads directly to
$\dot{\mathsf{E}}^{\top}=0$. This implies that the channel is independent
of the parameter, rendering parameter estimation impossible. Therefore,
for any metrology task where the channel carries information about
the parameter, the orthogonal component $\Lambda^{\perp}$ must be
nonzero.

\section{Proof of Asymptotically Tight Bound}

\label{sec:ultimate_bound}

Fix a differentiable minimal single-channel Kraus
representation $\{K_{i}\}^{r}_{i=1}$, where $r=\operatorname{rank}\mathsf{E}_{g}$
is the single-channel Choi rank. Retain $\alpha=\sum_{i}\dot{K}^{\dagger}_{i}\dot{K}_{i}$
and $\beta=\sum_{i}\dot{K}^{\dagger}_{i}K_{i}$. Set $P_{i}:=K_{i}\beta^{\dagger}$
and $R_{i}:=\dot{K}_{i}-P_{i}$. The completeness relation gives 
\begin{equation}
\sum_{i}R^{\dagger}_{i}K_{i}=0\;,\sum_{i}P^{\dagger}_{i}P_{i}=\beta\beta^{\dagger}\;,\sum_{i}P^{\dagger}_{i}R_{i}=0\;,\sum_{i}R^{\dagger}_{i}R_{i}=\alpha-\beta\beta^{\dagger}\succeq0.\label{eq:ultimate_local_identities}
\end{equation}

Let $C$, $P$, and $R$ collect the columns $|K^{\top}_{i}\rangle$,
$|P^{\top}_{i}\rangle$, and $|R^{\top}_{i}\rangle$, respectively.
For the product Kraus representation, define 
\begin{equation}
\mathbf{P}_{j}:=C^{\otimes(j-1)}\otimes P\otimes C^{\otimes(N-j)}\;,\mathbf{R}_{j}:=C^{\otimes(j-1)}\otimes R\otimes C^{\otimes(N-j)}.\label{eq:ultimate_embedded_PR}
\end{equation}
Its ensemble derivative is $\dot{\mathbf{C}}^{0}=\sum^{N}_{j=1}(\mathbf{P}_{j}+\mathbf{R}_{j})$.
This product representation is used only to construct an upper bound.

For a fixed $S\in\mathsf{Gen}$, denote the Hilbert--Schmidt
norm by $\|X\|^{2}_{2}=\operatorname{tr}(X^{\dagger}X)$. Evaluating
the inner minimization in Eq.~\eqref{eq:straqfi} of the main text
at $h=0$ gives 
\begin{equation}
\min_{h\in\mathbb{H}_{r^{N}}}\operatorname{tr}[\Omega_{g}(h)S]\le4\operatorname{tr}\left[S(\dot{\mathbf{C}}^{0}\dot{\mathbf{C}}^{0\dagger})^{\top}\right]=4\left\Vert \sqrt{S}^{\top}\sum^{N}_{j=1}(\mathbf{P}_{j}+\mathbf{R}_{j})\right\Vert ^{2}_{2}\le4\left[\left\Vert \sqrt{S}^{\top}\sum^{N}_{j=1}\mathbf{P}_{j}\right\Vert _{2}+\left\Vert \sqrt{S}^{\top}\sum^{N}_{j=1}\mathbf{R}_{j}\right\Vert _{2}\right]^{2}.\label{eq:ultimate_fixed_representation}
\end{equation}

To control the terms containing $R_{i}$, define
\begin{equation}
\Lambda_{R}:=4\sum_{i}|R^{\dagger}_{i}\rangle\langle K^{\dagger}_{i}|.\label{eq:ultimate_lambda_R}
\end{equation}
In the vectorization convention used throughout this Supplemental
Material, $(|X^{\top}\rangle\langle Y^{\top}|)^{\top}=|Y^{\dagger}\rangle\langle X^{\dagger}|$.
The first identity in Eq.~\eqref{eq:ultimate_local_identities} therefore
gives $\operatorname{tr}_{T^{\mathrm{o}}}\Lambda_{R}=4\sum_{i}R^{\dagger}_{i}K_{i}=0$.

For $j<k$, expanding the Kraus indices and linking
the remaining $N-2$ channel uses with their Choi operators gives
\begin{equation}
\operatorname{tr}(\mathbf{R}^{\dagger}_{k}S^{\top}\mathbf{R}_{j})=\operatorname{tr}\left[S(\mathbf{R}_{j}\mathbf{R}^{\dagger}_{k})^{\top}\right]=\frac{1}{16}\operatorname{tr}\left[(\Lambda^{\dagger}_{R,j}\otimes\Lambda_{R,k})S^{\mathrm{red}}_{jk}\right]=0.\label{eq:ultimate_cross_term_zero}
\end{equation}
Here $S^{\mathrm{red}}_{jk}$ is the reduced two-use process defined
in Eq.~\eqref{eq:reduced_two_use_process}. The operator $\Lambda_{R,j}$
denotes $\Lambda_{R}$ from Eq.~\eqref{eq:ultimate_lambda_R} acting
on the input and output spaces of channel use $j$, and $\Lambda_{R,k}$
is defined analogously. The two-use specialization of the projector
$\mathcal{Q}_{\mathsf{Gen}}$ in Eq.~\eqref{eq:qico_appendix} is
\begin{equation}
\mathcal{Q}_{\mathsf{Gen},2}:=(\mathcal{I}-\mathcal{D}_{T^{\mathrm{o}}_{j}}+\mathcal{D}_{T^{\mathrm{i}}_{j}T^{\mathrm{o}}_{j}})\otimes(\mathcal{I}-\mathcal{D}_{T^{\mathrm{o}}_{k}}+\mathcal{D}_{T^{\mathrm{i}}_{k}T^{\mathrm{o}}_{k}})-\mathcal{D}_{T^{\mathrm{i}}_{j}T^{\mathrm{o}}_{j}T^{\mathrm{i}}_{k}T^{\mathrm{o}}_{k}}.\label{eq:ultimate_QGen2}
\end{equation}
Both $\Lambda^{\dagger}_{R,j}$ and $\Lambda_{R,k}$ have zero partial
trace over their respective output spaces. Equation~\eqref{eq:charaoficq}
therefore gives $\Lambda^{\dagger}_{R,j}\otimes\Lambda_{R,k}\in\operatorname{im}\mathcal{Q}_{\mathsf{Gen},2}$.
Since $S^{\mathrm{red}}_{jk}$ is a valid two-use process, $S^{\mathrm{red}}_{jk}\in\ker\mathcal{Q}_{\mathsf{Gen},2}$.
The projector $\mathcal{Q}_{\mathsf{Gen},2}$ is self-adjoint under
the Hilbert--Schmidt trace pairing, so its image is orthogonal to
its kernel. This proves the last equality in Eq.~\eqref{eq:ultimate_cross_term_zero}.

Linking all channel uses except $j$ gives the valid
one-use process $S^{\mathrm{red}}_{j}=\rho_{j}\otimes\mathbb{I}_{T^{\mathrm{o}}_{j}}$,
where $\rho_{j}\succeq0$ and $\operatorname{tr}\rho_{j}=1$. Equation~\eqref{eq:ultimate_local_identities}
then gives 
\begin{equation}
\left\Vert \sqrt{S}^{\top}\mathbf{P}_{j}\right\Vert ^{2}_{2}=\operatorname{tr}\left[\rho_{j}\sum_{i}P^{\dagger}_{i}P_{i}\right]\le\|\beta\|^{2}.\label{eq:ultimate_P_bound}
\end{equation}
\begin{equation}
\left\Vert \sqrt{S}^{\top}\mathbf{R}_{j}\right\Vert ^{2}_{2}=\operatorname{tr}\left[\rho_{j}\sum_{i}R^{\dagger}_{i}R_{i}\right]\le\|\alpha-\beta\beta^{\dagger}\|.\label{eq:ultimate_R_bound}
\end{equation}

The Hilbert--Schmidt triangle inequality and Eq.~\eqref{eq:ultimate_P_bound}
imply 
\begin{equation}
\left\Vert \sqrt{S}^{\top}\sum^{N}_{j=1}\mathbf{P}_{j}\right\Vert _{2}\le\sum^{N}_{j=1}\left\Vert \sqrt{S}^{\top}\mathbf{P}_{j}\right\Vert _{2}\le N\|\beta\|.\label{eq:ultimate_P_sum}
\end{equation}
Equation~\eqref{eq:ultimate_cross_term_zero} eliminates all terms
with $j\neq k$. Equation~\eqref{eq:ultimate_R_bound} bounds the
remaining diagonal terms, giving 
\begin{equation}
\left\Vert \sqrt{S}^{\top}\sum^{N}_{j=1}\mathbf{R}_{j}\right\Vert ^{2}_{2}=\sum^{N}_{j,k=1}\operatorname{tr}(\mathbf{R}^{\dagger}_{k}S^{\top}\mathbf{R}_{j})=\sum^{N}_{j=1}\left\Vert \sqrt{S}^{\top}\mathbf{R}_{j}\right\Vert ^{2}_{2}\le N\|\alpha-\beta\beta^{\dagger}\|.\label{eq:ultimate_R_sum}
\end{equation}

Combining Eqs.~\eqref{eq:ultimate_fixed_representation},
\eqref{eq:ultimate_P_sum}, and \eqref{eq:ultimate_R_sum} gives,
for every valid $S$ and every single-channel Kraus representation,
\begin{equation}
\min_{h\in\mathbb{H}_{r^{N}}}\operatorname{tr}[\Omega_{g}(h)S]\le4\left[N\|\beta\|+\sqrt{N\|\alpha-\beta\beta^{\dagger}\|}\right]^{2}.\label{eq:ultimate_bound_supp}
\end{equation}
The right-hand side is independent of $S$. Maximizing over $S\in\mathsf{Gen}$
and minimizing over the single-channel Kraus representations proves
Eq.~\eqref{eq:bound3} of the main text.

If a Kraus representation with $\beta=0$ exists,
Eq.~\eqref{eq:ultimate_bound_supp} gives $\mathcal{F}^{\mathsf{Gen}}_{N}\le4N\min_{\{K_{i}\}:\beta=0}\|\alpha\|$.
Together with the known attainability by parallel strategies~\citep{zhou2021AsymptoticTheoryQuantum},
this fixes the same asymptotic linear coefficient.

When $\min_{\{K_{i}\}}\|\beta\|\neq0$, dividing
Eq.~\eqref{eq:ultimate_bound_supp} by $N^{2}$ and taking $N\to\infty$
gives 
\begin{equation}
\lim_{N\to\infty}\frac{\mathcal{F}^{\mathsf{Gen}}_{N}}{N^{2}}\le4\min_{\{K_{i}\}}\|\beta\|^{2}.\label{eq:ultimate_asymptotic_upper}
\end{equation}
Parallel strategies form a subset of general ICO strategies and attain
the same coefficient~\citep{zhou2021AsymptoticTheoryQuantum}. The
reverse inequality therefore follows, proving Eq.~\eqref{eq:general_asymptotic_equivalence}
of the main text and the QFI ratio stated below it whenever the common
coefficient is nonzero.

\section{Absence of ICO Advantage for Pauli Channels}

\label{sec:paulichannel}

We prove that for Pauli channels the optimal ICO strategy yields the
same SQL as parallel strategies, analytically confirming the numerical
conjecture of Ref.~\citep{mothe2024ReassessingAdvantageIndefinite}.
Consider 
\begin{equation}
\mathcal{E}_{\theta}(\rho)=\theta\rho+t\sum^{3}_{k=1}p_{k}\,\sigma_{k}\rho\sigma_{k},\;\theta\in(0,1),
\end{equation}
where $t:=1-\theta$ and $\{p_{k}\}$ is a probability distribution.
We claim 
\begin{equation}
\mathcal{F}^{\mathsf{Gen}}_{N}(\mathcal{E}_{\theta})=\frac{N}{\theta t}.\label{eq:pauli_sql}
\end{equation}

\textit{Proof.}---We verify the KKT conditions for the QFI optimization
\eqref{eq:straqfi}. Choose the ensemble matrix $C_{\theta}$ whose
columns are the vectorized Kraus operators: 
\begin{equation}
C_{\theta}=\bigl(\sqrt{\theta}\,|\mathbb{I}\rangle,\sqrt{tp_{1}}\,|\sigma^{\top}_{1}\rangle,\sqrt{tp_{2}}\,|\sigma^{\top}_{2}\rangle,\sqrt{tp_{3}}\,|\sigma^{\top}_{3}\rangle\bigr),\label{eq:pauensem}
\end{equation}
and propose the candidate $S^{\star}=2^{-N}\mathbb{I}_{T^{\mathrm{i}}_{\mathrm{tot}}T^{\mathrm{o}}_{\mathrm{tot}}}$
and $h^{\star}=0$ (recall $d=2$). Since $S^{\star}$ is a valid
parallel state, it is feasible for $\mathsf{Gen}$. \emph{QFI of the
candidate.} With $h=0$, the performance operator is $\Omega^{(N)}(0)=4\left[\dot{\mathbf{C}}^{0}_{\theta}(\dot{\mathbf{C}}^{0}_{\theta})^{\dagger}\right]^{\top}$
. Because $\mathcal{E}^{\otimes N}_{\theta}$ is a Pauli channel,
the off-diagonal contributions in $\dot{\mathbf{C}}^{0}_{\theta}$
vanish under the trace with the identity, giving 
\begin{equation}
\operatorname{tr}\!\bigl(\Omega^{(N)}(0)S^{\star}\bigr)=\frac{N}{\theta t}.
\end{equation}
Thus the candidate reproduces \eqref{eq:pauli_sql}.

\emph{KKT verification.} The Lagrangian stationarity with respect
to $h$ requires $\mathbf{C}^{0\dagger}_{\theta}S^{\star\top}\dot{\mathbf{C}}^{0}_{\theta}\propto\mathbf{C}^{0\dagger}_{\theta}\dot{\mathbf{C}}^{0}_{\theta}$
to be Hermitian. For the Pauli ensemble \eqref{eq:pauensem} one directly
checks that $C^{\dagger}_{\theta}\dot{C}_{\theta}$ is Hermitian.
Since $\mathbf{C}^{0}_{\theta}=C^{\otimes N}_{\theta}$, the tensor
product preserves this property, so $h^{\star}=0$ is stationary.
Because $S^{\star}$ is full-rank, complementary slackness forces
the dual variable $Y$ associated with $S\succeq0$ to vanish: $Y=0$.
With $Y=0$ and $h=0$, the stationarity with respect to $S$ reduces
to 
\begin{equation}
\Omega^{(N)}(0)-\mu\mathbb{I}=\mathcal{Q}^{\dagger}_{\mathsf{Gen}}(E),\label{eq:pauli_stationary}
\end{equation}
with $\mu=\operatorname{tr}(\Omega^{(N)}(0))/4^{N}$. Since $\mathcal{Q}_{\mathsf{Gen}}$
is a Hermitian projector, \eqref{eq:pauli_stationary} admits a solution
$E$ iff $X:=\Omega^{(N)}(0)-\mu\mathbb{I}$ lies in $\mathrm{im}\,\mathcal{Q}_{\mathsf{Gen}}$.
By construction $X$ is traceless. It remains to verify the local
constraint $\operatorname{tr}_{T^{\mathrm{o}}_{k}}(X)\propto\mathbb{I}_{T^{\mathrm{i}}_{k}}$
for each $k$. Using \eqref{eq:pauensem}, the single-channel operators
satisfy 
\begin{equation}
\operatorname{tr}_{T^{\mathrm{o}}}(C_{\theta}C^{\dagger}_{\theta})=\mathbb{I}_{T^{\mathrm{i}}},\quad\operatorname{tr}_{T^{\mathrm{o}}}(\dot{C}_{\theta}\dot{C}^{\dagger}_{\theta})\propto\mathbb{I}_{T^{\mathrm{i}}},\quad\operatorname{tr}_{T^{\mathrm{o}}}(C_{\theta}\dot{C}^{\dagger}_{\theta})\propto\mathbb{I}_{T^{\mathrm{i}}}.
\end{equation}
Because $\Omega^{(N)}(0)$ is a sum of tensor products of these three
objects, its partial trace over any output space $T^{\mathrm{o}}_{k}$
factorizes into $\mathbb{I}_{T^{\mathrm{i}}_{k}}\otimes B$ for some
remainder operator $B$. The identity term $\mu\mathbb{I}$ trivially
shares this property, hence $\operatorname{tr}_{T^{\mathrm{o}}_{k}}(X)\propto\mathbb{I}_{T^{\mathrm{i}}_{k}}$.
Therefore $X\in\mathrm{im}\,\mathcal{Q}_{\mathsf{Gen}}$, and we may
take $E=X$. The tuple $(S^{\star},h^{\star}=0,Y=0,E=X,\mu)$ satisfies
all KKT conditions, proving global optimality of the parallel strategy.
\hfill{}$\blacksquare$

\section{Iterative QFI Bound for QC-QC Strategies}

\label{sec:qcqc}

This section derives a finite-query iterative QFI bound for quantum
circuits with quantum control of causal order (QC-QC), which admit
a clear operational interpretation as generalized quantum circuits.
We extend the iterative method of Ref.~\citep{kurdzialek2023UsingAdaptivenessCausal}
and show that QC-QC strategies satisfy the same recurrence as causal-superposition
strategies.

\subsection{Formalization of QC-QC Strategies}

We first provide a formal description of QC-QC strategies to lay the
foundation for the subsequent QFI analysis. Our description follows
the constructive framework established by Wechs et al.~\citep{wechs2021QuantumCircuitsClassical},
which represents a process involving $N$ uses of the external channel
$\mathcal{E}_{g}$ as an overall evolution constructed from the chain
multiplication of a series of internal isometries $\tilde{V}_{n}$
and controlled applications of external channels $\tilde{A}_{n}$.
To accurately describe the physical picture, we distinguish between
time steps (indexed by $n=1,\dots,N$) and external channel indices
(indexed by $k\in\mathcal{N}:=\{1,\dots,N\}$). A QC-QC circuit introduces
a control system whose basis vector at the $n$-th time step is $|\mathcal{K}_{n-1},k_{n}\rangle^{C_{n}}$.
This basis vector records the set $\mathcal{K}_{n-1}$ of $n-1$ channels
that have already acted before this time step, as well as the index
$k_{n}$ of the channel acting at the current time step. Under this
control, the external channel $\mathcal{E}_{k_{n}}$ (with Kraus operators
$K_{\nu_{k_{n}}}$) is applied at the $n$-th time step. This controlled
operation can be described by the following operator: 
\begin{equation}
\tilde{A}_{n}(\vec{\nu})=\sum_{\mathcal{K}_{n-1},k_{n}}K_{\nu_{k_{n}}}\otimes|\mathcal{K}_{n-1},k_{n}\rangle^{C^{\prime}_{n}}\langle\mathcal{K}_{n-1},k_{n}|^{C_{n}},
\end{equation}
where $C_{n}$ and $C_{n}'$ label the control system before and after
the channel action, respectively. These controlled external channel
applications are coherently connected by a series of internal isometries
$\tilde{V}_{n}$. These internal operators not only control the execution
order of the target system but also update the state of the control
system, introducing ancillary systems when necessary to ensure the
isometry of the operations. The initial operator $\tilde{V}_{1}$
is responsible for preparing the input state from the global past
(P) and coherently distributing it to the first possible channel to
act: 
\begin{equation}
\tilde{V}_{1}=\sum_{k_{1}\in\mathcal{N}}V^{\to k_{1}}_{\emptyset,\emptyset}\otimes|\emptyset,k_{1}\rangle^{C_{1}}.
\end{equation}
The intermediate operator $\tilde{V}_{n+1}$ (for $1\le n<N$) connects
the output of the $n$-th step to the input of the $(n+1)$-th step:
\begin{equation}
\tilde{V}_{n+1}=\sum_{\substack{\mathcal{K}_{n-1},k_{n}\\
k_{n+1}\notin\mathcal{K}_{n}
}
}V^{\to k_{n+1}}_{\mathcal{K}_{n-1},k_{n}}\otimes|\mathcal{K}_{n},k_{n+1}\rangle^{C_{n+1}}\langle\mathcal{K}_{n-1},k_{n}|^{C_{n}'},
\end{equation}
where $\mathcal{K}_{n}=\mathcal{K}_{n-1}\cup\{k_{n}\}$. Finally,
the terminating operator $\tilde{V}_{N+1}$ collects the outputs of
all paths to the global future (F): 
\begin{equation}
\tilde{V}_{N+1}=\sum_{k_{N}\in\mathcal{N}}V^{\to F}_{\mathcal{N}\setminus\{k_{N}\},k_{N}}\otimes\langle\mathcal{N}\setminus\{k_{N}\},k_{N}|^{C_{N}'}.
\end{equation}
By multiplying the operators defined above, we obtain the global Kraus
operator describing the complete process from $\mathcal{H}^{P}$ to
$\mathcal{H}^{F}$: 
\begin{equation}
\mathbf{K}^{(N)}_{\vec{\nu}}=\tilde{V}_{N+1}\tilde{A}_{N}(\vec{\nu})\tilde{V}_{N}\cdots\tilde{V}_{2}\tilde{A}_{1}(\vec{\nu})\tilde{V}_{1}.
\end{equation}
This operator depends on the Kraus operator indices $\vec{\nu}=(\nu_{1},\dots,\nu_{N})$
of all $N$ external channels. Although this expression is formally
similar to the Kraus operator for adaptive (AD) strategies, the key
difference lies in the fact that the summation in the internal operators
$\tilde{V}_{n}$ represents a coherent superposition of different
causal paths rather than a classical probabilistic choice, manifesting
its quantum nature. After contracting the control system, we can write
the total Kraus operator as: 
\begin{equation}
\mathbf{K}^{(N)}_{\vec{\nu}}=\sum_{\pi=(k_{1},\dots,k_{N})\in S_{N}}V^{\to F}_{\mathcal{K}_{N-1},k_{N}}K_{\nu_{k_{N}}}\cdots K_{\nu_{k_{1}}}V^{\to k_{1}}_{\emptyset,\emptyset}.
\end{equation}
We can also define an operator $\mathbf{K}^{(n)}_{\vec{\nu}}$ describing
the evolution of the first $n$ steps, which satisfies the following
iterative relation: 
\begin{equation}
\mathbf{K}^{(n)}_{\vec{\nu}}=\tilde{V}_{n+1}\tilde{A}_{n}(\vec{\nu})\mathbf{K}^{(n-1)}_{\vec{\nu}},\label{eq:qcqc_kraus_iterative}
\end{equation}
where the initial condition of the recursion is defined as $\mathbf{K}^{(0)}_{\vec{\nu}}=\tilde{V}_{1}$.
This iterative form is central to the subsequent derivation of the
iterative bound. Before proceeding with the iterative derivation,
we first prove an important lemma that reveals the normalization property
of the intermediate operator $\mathbf{K}^{(n)}_{\vec{\nu}}$. Unlike
the complete $N$-step process, an ``incomplete'' process containing
only the first $n$ steps is not trace-preserving. Specifically, we
need to calculate $\sum_{\vec{\nu}}(\mathbf{K}^{(n)}_{\vec{\nu}})^{\dagger}\mathbf{K}^{(n)}_{\vec{\nu}}$.
This summation iterates over all Kraus operator indices for all $N$
channels. Since the construction of $\mathbf{K}^{(n)}_{\vec{\nu}}$
only involves channels on the first $n$ paths, summing over the Kraus
indices of the $N-n$ channels that have not yet occurred will produce
an additional coefficient.

\textbf{Lemma 2.} For any $n\in\{0,\dots,N\}$, the intermediate Kraus
operators satisfy the following relation:

\begin{equation}
\sum_{\vec{\nu}}(\mathbf{K}^{(n)}_{\vec{\nu}})^{\dagger}\mathbf{K}^{(n)}_{\vec{\nu}}=r^{N-n}\mathbb{I}_{\mathcal{H}^{P}},
\end{equation}
where $r=\operatorname{rank}\mathsf{E}_{g}$ is the minimal number
of Kraus operators for a single channel.

\textit{Proof.} Let $\mathcal{P}^{(n)}=\sum_{\vec{\nu}}(\mathbf{K}^{(n)}_{\vec{\nu}})^{\dagger}\mathbf{K}^{(n)}_{\vec{\nu}}$
. For $n=0$, the isometry of $\tilde{V}_{1}$ and the sums over all
unused Kraus indices give $\mathcal{P}^{(0)}=r^{N}\mathbb{I}_{\mathcal{H}^{P}}$.
For $n=N$, the terminal isometry $\tilde{V}_{N+1}$ and trace preservation
of the complete QC-QC channel give $\mathcal{P}^{(N)}=\mathbb{I}_{\mathcal{H}^{P}}$.
It remains to consider $1\le n<N$. Using the iterative construction
of $\mathbf{K}^{(n)}_{\vec{\nu}}$, we can proceed with the following
derivation. First, we expand $\mathbf{K}^{(n)}_{\vec{\nu}}$ as a
superposition of control system basis vectors and utilize the orthogonality
of the control states $|\mathcal{K}_{n},k_{n+1}\rangle$ and $|\mathcal{Q}_{n},q_{n+1}\rangle$
to eliminate cross terms between different causal paths, retaining
only the diagonal terms (i.e., terms where $\mathcal{K}_{n}=\mathcal{Q}_{n}$
and $k_{n+1}=q_{n+1}$):

\begin{align}
\mathcal{P}^{(n)} & =\sum_{\vec{\nu}}\left(\sum_{\vec{k}_{n+1}}V^{\to k_{n+1}}_{\mathcal{K}_{n-1},k_{n}}K_{\nu_{k_{n}}}\cdots K_{\nu_{k_{1}}}V^{\to k_{1}}_{\emptyset,\emptyset}\otimes|\mathcal{K}_{n},k_{n+1}\rangle\right)^{\dagger}\nonumber \\
 & \quad\times\left(\sum_{\vec{q}_{n+1}}V^{\to q_{n+1}}_{\mathcal{Q}_{n-1},q_{n}}K_{\nu_{q_{n}}}\cdots K_{\nu_{q_{1}}}V^{\to q_{1}}_{\emptyset,\emptyset}\otimes|\mathcal{Q}_{n},q_{n+1}\rangle\right)\nonumber \\
 & =\sum_{\vec{\nu}}\sum_{\vec{k}_{n+1},\vec{q}_{n+1}}\left(V^{\to k_{n+1}}_{\mathcal{K}_{n-1},k_{n}}K_{\nu_{k_{n}}}\cdots K_{\nu_{k_{1}}}V^{\to k_{1}}_{\emptyset,\emptyset}\right)^{\dagger}\nonumber \\
 & \quad\times\left(V^{\to q_{n+1}}_{\mathcal{Q}_{n-1},q_{n}}K_{\nu_{q_{n}}}\cdots K_{\nu_{q_{1}}}V^{\to q_{1}}_{\emptyset,\emptyset}\right)\cdot\delta_{\mathcal{K}_{n},\mathcal{Q}_{n}}\delta_{k_{n+1},q_{n+1}}.
\end{align}
Next, we use the isometric property of the internal routing operator
$\tilde{V}_{n+1}$ to sum over $k_{n+1}$. According to the orthonormality
condition $\sum_{k_{n+1}}(V^{\to k_{n+1}}_{\mathcal{K}_{n-1},k_{n}})^{\dagger}V^{\to k_{n+1}}_{\mathcal{Q}_{n-1},q_{n}}=\delta_{\mathcal{K}_{n-1},\mathcal{Q}_{n-1}}\delta_{k_{n},q_{n}}\mathbb{I}_{\text{sys}}$,
this further constrains the path such that $k_{n}=q_{n}$:
\begin{equation}
\mathcal{P}^{(n)}=\sum_{\vec{\nu}}\sum_{\substack{\vec{k}_{n},\vec{q}_{n}\\
\mathcal{K}_{n}=\mathcal{Q}_{n}
}
}\left(K_{\nu_{k_{n}}}\cdots K_{\nu_{k_{1}}}V^{\to k_{1}}_{\emptyset,\emptyset}\right)^{\dagger}\left(\delta_{\mathcal{K}_{n-1},\mathcal{Q}_{n-1}}\delta_{k_{n},q_{n}}\mathbb{I}_{\text{sys}}\right)\left(K_{\nu_{q_{n}}}\cdots K_{\nu_{q_{1}}}V^{\to q_{1}}_{\emptyset,\emptyset}\right).
\end{equation}
 We use $\delta_{k_{n},q_{n}}$ to replace $q_{n}$ with $k_{n}$,
and sum over the Kraus operator index $\nu_{k_{n}}$ of the $n$-th
channel. Using the trace-preserving property of the channel $\sum_{\nu_{k_{n}}}K^{\dagger}_{\nu_{k_{n}}}K_{\nu_{k_{n}}}=\mathbb{I}_{\text{sys}}$,
the operation at the $n$-th layer is eliminated: 
\begin{align}
\mathcal{P}^{(n)} & =\sum_{\vec{\nu}_{\setminus k_{n}}}\sum_{\substack{\vec{k}_{n-1},\vec{q}_{n-1}\\
\mathcal{K}_{n-1}=\mathcal{Q}_{n-1}
}
}\left(V^{\to k_{n}}_{\mathcal{K}_{n-2},k_{n-1}}K_{\nu_{k_{n-1}}}\cdots K_{\nu_{k_{1}}}V^{\to k_{1}}_{\emptyset,\emptyset}\right)^{\dagger}\left(V^{\to k_{n}}_{\mathcal{Q}_{n-2},q_{n-1}}K_{\nu_{q_{n-1}}}\cdots K_{\nu_{q_{1}}}V^{\to q_{1}}_{\emptyset,\emptyset}\right).
\end{align}
The form of the expression has now returned to the $(n-1)$-th layer.
We repeat the above steps: first sum over $k_{n}$ using the isometry
of $\tilde{V}_{n}$, then sum over $\nu_{k_{n-1}}$ using the trace-preserving
relation $\sum_{\nu_{k_{n-1}}}K^{\dagger}_{\nu_{k_{n-1}}}K_{\nu_{k_{n-1}}}=\mathbb{I}_{\mathrm{sys}}$
. Summing over $k_{n}$: 
\begin{align}
\mathcal{P}^{(n)} & =\sum_{\vec{\nu}_{\setminus k_{n}}}\sum_{\substack{\vec{k}_{n-1},\vec{q}_{n-1}\\
\mathcal{K}_{n-1}=\mathcal{Q}_{n-1}
}
}\left(K_{\nu_{k_{n-1}}}\cdots K_{\nu_{k_{1}}}V^{\to k_{1}}_{\emptyset,\emptyset}\right)^{\dagger}\left(\sum_{k_{n}}(V^{\to k_{n}}_{\mathcal{K}_{n-2},k_{n-1}})^{\dagger}V^{\to k_{n}}_{\mathcal{Q}_{n-2},q_{n-1}}\right)\left(K_{\nu_{q_{n-1}}}\cdots K_{\nu_{q_{1}}}V^{\to q_{1}}_{\emptyset,\emptyset}\right)\nonumber \\
 & =\sum_{\vec{\nu}_{\setminus k_{n}}}\sum_{\substack{\vec{k}_{n-1},\vec{q}_{n-1}\\
\mathcal{K}_{n-1}=\mathcal{Q}_{n-1}
}
}\left(K_{\nu_{k_{n-1}}}\cdots K_{\nu_{k_{1}}}V^{\to k_{1}}_{\emptyset,\emptyset}\right)^{\dagger}\left(\delta_{\mathcal{K}_{n-2},\mathcal{Q}_{n-2}}\delta_{k_{n-1},q_{n-1}}\mathbb{I}_{\text{sys}}\right)\left(K_{\nu_{q_{n-1}}}\cdots K_{\nu_{q_{1}}}V^{\to q_{1}}_{\emptyset,\emptyset}\right).
\end{align}
Summing over $\nu_{k_{n-1}}$: 
\begin{align}
\mathcal{P}^{(n)} & =\sum_{\vec{\nu}_{\setminus\{k_{n},k_{n-1}\}}}\sum_{\substack{\vec{k}_{n-2},\vec{q}_{n-2}\\
\mathcal{K}_{n-2}=\mathcal{Q}_{n-2}
}
}\left(V^{\to k_{n-1}}_{\mathcal{K}_{n-3},k_{n-2}}\cdots K_{\nu_{k_{1}}}V^{\to k_{1}}_{\emptyset,\emptyset}\right)^{\dagger}\left(V^{\to k_{n-1}}_{\mathcal{Q}_{n-3},q_{n-2}}\cdots K_{\nu_{q_{1}}}V^{\to q_{1}}_{\emptyset,\emptyset}\right).
\end{align}
Analogously, we peel off layers inwardly until the first layer. At
the first layer, we sum over $k_{2}$ to eliminate $\tilde{V}_{2}$
, and sum over $\nu_{k_{1}}$ using the same trace-preserving relation:
\begin{align}
\mathcal{P}^{(n)} & =\sum_{\vec{\nu}_{\text{unused}}}\sum_{k_{1},q_{1}}(V^{\to k_{1}}_{\emptyset,\emptyset})^{\dagger}K^{\dagger}_{\nu_{k_{1}}}\left(\sum_{k_{2}}(V^{\to k_{2}}_{\emptyset,k_{1}})^{\dagger}V^{\to k_{2}}_{\emptyset,q_{1}}\right)K_{\nu_{q_{1}}}V^{\to q_{1}}_{\emptyset,\emptyset}\nonumber \\
 & =\sum_{\vec{\nu}_{\text{unused}}}\sum_{k_{1}}(V^{\to k_{1}}_{\emptyset,\emptyset})^{\dagger}\left(\sum_{\nu_{k_{1}}}K^{\dagger}_{\nu_{k_{1}}}K_{\nu_{k_{1}}}\right)V^{\to k_{1}}_{\emptyset,\emptyset}.
\end{align}
Finally, using the normalization condition of the initial operator
$\sum_{k_{1}}(V^{\to k_{1}}_{\emptyset,\emptyset})^{\dagger}V^{\to k_{1}}_{\emptyset,\emptyset}=\mathbb{I}_{\mathcal{H}^{P}}$,
and summing over the Kraus indices $\vec{\nu}_{\text{unused}}$ of
the $N-n$ channels not participating in the process (which contributes
a factor of $r^{N-n}$), we obtain the final result: 
\begin{align}
\mathcal{P}^{(n)} & =\sum_{\vec{\nu}_{\text{unused}}}\mathbb{I}_{\mathcal{H}^{P}}=r^{N-n}\mathbb{I}_{\mathcal{H}^{P}}.
\end{align}

This completes the proof of the lemma. \hfill{}$\blacksquare$

\subsection{Derivation of Single-Step Iterative Bound for QFI}

For the parameter estimation problem, the upper bound on the maximum
achievable QFI after $N$ channel calls is determined by the derivative
properties of the global Kraus operator with respect to the parameter:

\begin{equation}
\mathcal{F}_{N}\le4\min_{\{K_{i}\}}\left\Vert \sum_{\vec{\nu}}(\dot{\mathbf{K}}^{(N)}_{\vec{\nu}})^{\dagger}\dot{\mathbf{K}}^{(N)}_{\vec{\nu}}\right\Vert .
\end{equation}

To analyze the asymptotic behavior of this quantity, we define the
intermediate quantity $\alpha^{(n)}$ at the $n$-th step as:

\begin{equation}
\alpha^{(n)}:=\sum_{\vec{\nu}}(\dot{\mathbf{K}}^{(n)}_{\vec{\nu}})^{\dagger}\dot{\mathbf{K}}^{(n)}_{\vec{\nu}}.
\end{equation}
The summation $\sum_{\vec{\nu}}$ still iterates over the Kraus indices
of all $N$ channels, even though $\mathbf{K}^{(n)}_{\vec{\nu}}$
only depends on the channels involved in the first $n$ steps. This
results in $\alpha^{(n)}$ containing an additional coefficient from
the inactive channels. To eliminate this effect and obtain an iterative
relation with clear physical meaning, we introduce the normalized
operator $\tilde{\alpha}^{(n)}:=r^{-(N-n)}\alpha^{(n)}$. We will
prove that the norm of $\tilde{\alpha}^{(n)}$ satisfies the following
single-step iterative relation.

\textbf{Theorem A.} For QC-QC strategies, the norm of the normalized
operator satisfies the following single-step iterative relation: 
\begin{equation}
\|\tilde{\alpha}^{(n)}\|\le\|\tilde{\alpha}^{(n-1)}\|+\|\alpha\|+2\|\beta\|\sqrt{\|\tilde{\alpha}^{(n-1)}\|}.\label{eq:iterative_bound}
\end{equation}

\textit{Proof.} First, we expand $\dot{\mathbf{K}}^{(n)}_{\vec{\nu}}$
using the iterative definition of the Kraus operator~\eqref{eq:qcqc_kraus_iterative}
and the Leibniz rule, yielding $\dot{\mathbf{K}}^{(n)}_{\vec{\nu}}=\tilde{V}_{n+1}(\dot{\tilde{A}}_{n}(\vec{\nu})\mathbf{K}^{(n-1)}_{\vec{\nu}}+\tilde{A}_{n}(\vec{\nu})\dot{\mathbf{K}}^{(n-1)}_{\vec{\nu}})$.
Using the isometry of the internal operator $\tilde{V}_{n+1}$, $\alpha^{(n)}$
can be expressed as the sum of three terms:

\begin{align}
\alpha^{(n)} & =\sum_{\vec{\nu}}\left(\dot{\tilde{A}}_{n}(\vec{\nu})\mathbf{K}^{(n-1)}_{\vec{\nu}}+\tilde{A}_{n}(\vec{\nu})\dot{\mathbf{K}}^{(n-1)}_{\vec{\nu}}\right)^{\dagger}\left(\dot{\tilde{A}}_{n}(\vec{\nu})\mathbf{K}^{(n-1)}_{\vec{\nu}}+\tilde{A}_{n}(\vec{\nu})\dot{\mathbf{K}}^{(n-1)}_{\vec{\nu}}\right)\nonumber \\
 & =Q_{1}+Q_{2}+Q_{3},
\end{align}
where 
\begin{align}
Q_{1} & :=\sum_{\vec{\nu}}(\mathbf{K}^{(n-1)}_{\vec{\nu}})^{\dagger}\dot{\tilde{A}}^{\dagger}_{n}(\vec{\nu})\dot{\tilde{A}}_{n}(\vec{\nu})\mathbf{K}^{(n-1)}_{\vec{\nu}},\label{eq:localQ}\\
Q_{2} & :=\sum_{\vec{\nu}}(\dot{\mathbf{K}}^{(n-1)}_{\vec{\nu}})^{\dagger}\tilde{A}^{\dagger}_{n}(\vec{\nu})\tilde{A}_{n}(\vec{\nu})\dot{\mathbf{K}}^{(n-1)}_{\vec{\nu}},\label{eq:historyQ}\\
Q_{3} & :=\sum_{\vec{\nu}}\left((\mathbf{K}^{(n-1)}_{\vec{\nu}})^{\dagger}\dot{\tilde{A}}^{\dagger}_{n}(\vec{\nu})\tilde{A}_{n}(\vec{\nu})\dot{\mathbf{K}}^{(n-1)}_{\vec{\nu}}+\text{h.c.}\right).\label{eq:crossQ}
\end{align}

These three terms correspond to the local contribution, the accumulation
contribution, and their cross term, respectively. We need to analyze
these three parts term by term. For $Q_{1}$ (the local term,~\eqref{eq:localQ}),
we first sum over the Kraus index of the channel selected at the $n$-th
step. Using the definition of the controlled external channel $\tilde{A}_{n}(\vec{\nu})$,
we have

\begin{equation}
\begin{aligned}\sum_{\nu_{k_{n}}}\dot{\tilde{A}}^{\dagger}_{n}\dot{\tilde{A}}_{n} & =\sum_{\mathcal{K}_{n-1},k_{n}}\left(\sum_{\nu_{k_{n}}}\dot{K}^{\dagger}_{\nu_{k_{n}}}\dot{K}_{\nu_{k_{n}}}\right)\otimes\Pi^{C_{n}}_{\mathcal{K}_{n-1},k_{n}}\\
 & =\alpha\otimes\mathbb{I}^{C_{n}},
\end{aligned}
\end{equation}
where $\Pi^{C_{n}}_{\mathcal{K}_{n-1},k_{n}}=|\mathcal{K}_{n-1},k_{n}\rangle\langle\mathcal{K}_{n-1},k_{n}|^{C_{n}}$
is the projection operator on the control system.

\[
\]
Substituting this back into the expression for $Q_{1}$, performing
this current-index sum inside $Q_{1}$, and noting that $\mathbf{K}^{(n-1)}_{\vec{\nu}}$
does not depend on $\nu_{k_{n}}$, we can write the local term as:
\begin{equation}
Q_{1}=\sum_{\vec{\nu}_{\setminus k_{n}}}(\mathbf{K}^{(n-1)}_{\vec{\nu}})^{\dagger}(\alpha\otimes\mathbb{I}^{C_{n}})\mathbf{K}^{(n-1)}_{\vec{\nu}}.
\end{equation}
Here, we apply the operator norm inequality 
\begin{equation}
\left\Vert \sum_{k}L^{\dagger}_{k}AM_{k}\right\Vert \le\left\Vert \sum_{k}L^{\dagger}_{k}L_{k}\right\Vert ^{1/2}\|A\|\left\Vert \sum_{k}M^{\dagger}_{k}M_{k}\right\Vert ^{1/2}.\label{eq:ineq}
\end{equation}
Let $L_{k}=M_{k}=\mathbf{K}^{(n-1)}_{\vec{\nu}}$, where $k$ corresponds
to $\vec{\nu}_{\setminus k_{n}}$, and $A=\alpha\otimes\mathbb{I}^{C_{n}}$.
At this point, the summation $\sum_{\vec{\nu}_{\setminus k_{n}}}(\mathbf{K}^{(n-1)}_{\vec{\nu}})^{\dagger}\mathbf{K}^{(n-1)}_{\vec{\nu}}$
is obtained from Lemma 2 at step $n-1$ by removing the sum over $\nu_{k_{n}}$.
Since $\mathbf{K}^{(n-1)}_{\vec{\nu}}$ is independent of $\nu_{k_{n}}$,
this removes a factor of $r$, and hence $\|\sum_{\vec{\nu}_{\setminus k_{n}}}(\mathbf{K}^{(n-1)}_{\vec{\nu}})^{\dagger}\mathbf{K}^{(n-1)}_{\vec{\nu}}\|=r^{N-n}$.
Therefore, we obtain:

\begin{equation}
\|Q_{1}\|\le r^{N-n}\|\alpha\|.
\end{equation}

For $Q_{2}$ (the accumulation term,~\eqref{eq:historyQ}), trace
preservation of the single-channel Kraus representation gives 
\begin{equation}
\begin{aligned}\sum_{\nu_{k_{n}}}\tilde{A}^{\dagger}_{n}\tilde{A}_{n} & =\sum_{\mathcal{K}_{n-1},k_{n}}\left(\sum_{\nu_{k_{n}}}K^{\dagger}_{\nu_{k_{n}}}K_{\nu_{k_{n}}}\right)\otimes\Pi^{C_{n}}_{\mathcal{K}_{n-1},k_{n}}\\
 & =\mathbb{I}_{\mathrm{sys}}\otimes\mathbb{I}^{C_{n}}.
\end{aligned}
\end{equation}
The accumulation term becomes: 
\begin{equation}
Q_{2}=\sum_{\vec{\nu}_{\setminus k_{n}}}(\dot{\mathbf{K}}^{(n-1)}_{\vec{\nu}})^{\dagger}\dot{\mathbf{K}}^{(n-1)}_{\vec{\nu}}.
\end{equation}
Noting that the definition of $\alpha^{(n-1)}$ includes summation
over all $N$ channels (including $\nu_{k_{n}}$), the above expression
is actually equal to $\frac{1}{r}\alpha^{(n-1)}$. Taking the operator
norm, we get $\|Q_{2}\|=\frac{1}{r}\|\alpha^{(n-1)}\|$. For $Q_{3}$
(the cross term,~\eqref{eq:crossQ}), the intermediate operator involves
$\sum_{\nu_{k_{n}}}\dot{\tilde{A}}^{\dagger}_{n}\tilde{A}_{n}$. With
$\beta=\sum_{i}\dot{K}^{\dagger}_{i}K_{i}$, this equals $\beta\otimes\mathbb{I}^{C_{n}}$
. Thus, the cross term can be written as:

\begin{equation}
Q_{3}=\sum_{\vec{\nu}_{\setminus k_{n}}}(\mathbf{K}^{(n-1)}_{\vec{\nu}})^{\dagger}(\beta\otimes\mathbb{I}^{C_{n}})\dot{\mathbf{K}}^{(n-1)}_{\vec{\nu}}+\text{h.c.}
\end{equation}
Applying the operator norm inequality~\eqref{eq:ineq} again with
$L_{k}=\mathbf{K}^{(n-1)}_{\vec{\nu}}$, $M_{k}=\dot{\mathbf{K}}^{(n-1)}_{\vec{\nu}}$
, and $A=\beta\otimes\mathbb{I}^{C_{n}}$, we have:

\begin{equation}
\|Q_{3}\|\le2\|\beta\|\sqrt{\left\Vert \sum_{\vec{\nu}_{\setminus k_{n}}}(\mathbf{K}^{(n-1)}_{\vec{\nu}})^{\dagger}\mathbf{K}^{(n-1)}_{\vec{\nu}}\right\Vert \cdot\left\Vert \sum_{\vec{\nu}_{\setminus k_{n}}}(\dot{\mathbf{K}}^{(n-1)}_{\vec{\nu}})^{\dagger}\dot{\mathbf{K}}^{(n-1)}_{\vec{\nu}}\right\Vert }.
\end{equation}
Substituting the results from the previous two steps, the value inside
the first square root is $r^{N-n}$, and the value inside the second
square root is $\frac{1}{r}\|\alpha^{(n-1)}\|$. This gives the upper
bound 
\begin{equation}
\|Q_{3}\|\le2\|\beta\|\sqrt{r^{N-n}\cdot\frac{1}{r}\|\alpha^{(n-1)}\|}.
\end{equation}
Combining the norm upper bounds for the three terms above, we obtain
the inequality for $\|\alpha^{(n)}\|$: 
\begin{equation}
\|\alpha^{(n)}\|\le r^{N-n}\|\alpha\|+\frac{1}{r}\|\alpha^{(n-1)}\|+2\|\beta\|\sqrt{r^{N-n-1}\|\alpha^{(n-1)}\|}.
\end{equation}

Using the renormalization definition $\tilde{\alpha}^{(n)}=r^{-(N-n)}\alpha^{(n)}$,
we have $\|\alpha^{(n)}\|=r^{N-n}\|\tilde{\alpha}^{(n)}\|$. The terms
on the right-hand side of the above equation can be rewritten as:
\begin{equation}
\|Q_{1}\|=r^{N-n}\|\alpha\|.
\end{equation}
\begin{equation}
\|Q_{2}\|=\frac{1}{r}\cdot r^{N-(n-1)}\|\tilde{\alpha}^{(n-1)}\|=r^{N-n}\|\tilde{\alpha}^{(n-1)}\|.
\end{equation}
\begin{equation}
\begin{aligned}\|Q_{3}\| & \le2\|\beta\|\sqrt{r^{N-n-1}\cdot r^{N-(n-1)}\|\tilde{\alpha}^{(n-1)}\|}\\
 & =r^{N-n}\cdot2\|\beta\|\sqrt{\|\tilde{\alpha}^{(n-1)}\|}.
\end{aligned}
\end{equation}
Canceling the common factor $r^{N-n}$ from both sides yields the
iterative relation in Theorem A: 
\begin{equation}
\|\tilde{\alpha}^{(n)}\|\le\|\tilde{\alpha}^{(n-1)}\|+\|\alpha\|+2\|\beta\|\sqrt{\|\tilde{\alpha}^{(n-1)}\|}.
\end{equation}
\hfill{}$\blacksquare$

Starting from $\tilde{\alpha}^{(0)}=0$, iteration
of Eq.~\eqref{eq:iterative_bound} gives a computable finite-query
QFI bound that is independent of the internal routing isometries.
Its recurrence is identical to that for causal-superposition strategies
in Ref.~\citep{kurdzialek2023UsingAdaptivenessCausal}. Hence QC-QC
and causal-superposition strategies share the same iterative upper-bound
characterization. Whenever the common leading coefficient is nonzero,
its asymptotic consequence is 
\begin{equation}
\lim_{N\to\infty}\mathcal{F}^{\mathsf{QC\text{-}QC}}_{N}/\mathcal{F}^{\mathsf{Para}}_{N}=1.
\end{equation}
 The asymptotic scaling coefficient obtained from this iterative bound
also agrees with the coefficient fixed for general ICO by the AT bound
in Eq.~\eqref{eq:bound3} of the main text.

\section{Additional Numerical Parameter Scans}

\label{sec:additional_numerical_scans}

\begin{figure}[ht]
\centering \includegraphics[width=0.95\columnwidth]{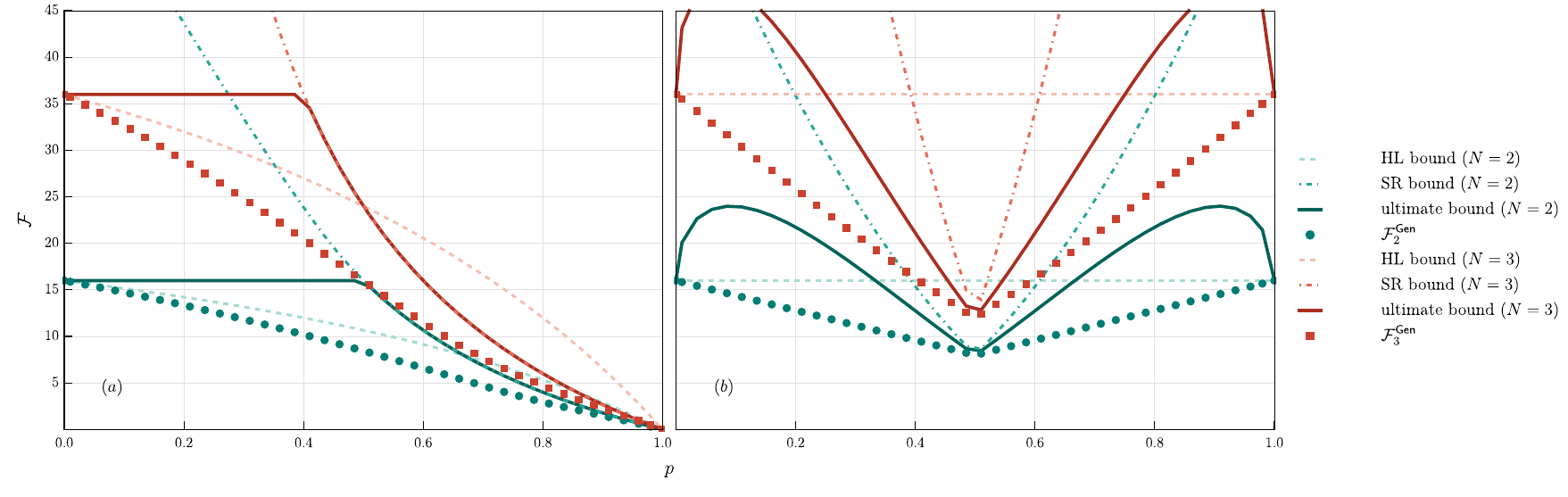}
\caption{Comparison of the numerically optimized general-ICO QFI $\mathcal{F}^{\mathsf{Gen}}_{N}$
with the universal HL, structurally refined (SR), and AT bounds versus
$p$ for $N=2,3$. Curves show the bounds, and unconnected symbols
show values obtained via semidefinite programming (SDP). (a) Generalized
amplitude damping (GAD) channel with $g=1/2$ and $\theta=\pi/4$.
(b) Perpendicular dephasing channel with $g=1/2$.}
\label{fig:numerical_parameter_scans}
\end{figure}

For completeness, Fig.~\ref{fig:numerical_parameter_scans}
compares the numerically optimized general-ICO QFI with the HL, SR,
and AT bounds as the channel noise parameter is varied for the two
channel families considered in the main text.

These parameter scans show that the representative
parameter choices in the main text are not exceptional. Over the GAD
noise range shown, the Hamiltonian-in-Kraus-span condition continues
to hold, so the optimized QFI remains in the SQL regime. The SR and
AT bounds follow the variation of the finite-query QFI more closely
than the universal HL bound. For perpendicular dephasing, $\min_{\{K_{i}\}}|\beta|\neq0$
throughout the scanned range, so the optimized QFI remain in the HL
regime. The noise parameter therefore changes the finite-$N$ QFI
and the tightness of these bounds, but does not alter the scaling
regime of either channel family.
\end{document}